\documentclass[twocolumn,
               superscriptaddress,
               floatfix,
               longbibliography, 
               showkeys,apl]{revtex4-2}

\usepackage{graphicx} 
\usepackage{bm}	
\usepackage{color}                     
\usepackage{xcolor}
\usepackage{epsfig}
\usepackage{amsmath} 
\usepackage{amssymb} 
\usepackage{longtable} 
\usepackage{float}
\usepackage{mathtools}
\usepackage{xparse}
\usepackage{hyperref}
\usepackage{times}
\usepackage[normalem]{ulem}
\usepackage{sidecap}
\usepackage{appendix} 
\sidecaptionvpos{figure}{t}
\usepackage{comment}

\newcommand{\zinit}{FLIP}
\newcommand{\params}{\pmb{\theta}} % Parameters of the PQC
\newcommand{\init}{\pmb{\theta}^{(0)}} % Initial parameter
\newcommand{\inittask}{\pmb{\theta}_\taskidx^{(0)}} % Initial parameters
\newcommand{\final}{\pmb{\theta}_\taskidx^{(s)}} % after k steps of optimization
\newcommand{\encoding}{\pmb{h}_k} % encoding of the p-th parameter
\newcommand{\encodingtask}{\pmb{h}^\taskidx_k} % encoding of the p-th parameter
\newcommand{\encodingtaskbis}{\pmb{h}^{\taskidx^\prime}_k} % encoding of the p-th parameter
\newcommand{\encodingbistask}{\pmb{h}^\taskidx_{k^\prime}} % encoding of the p-th parameter

\newcommand{\ansatz}{\mathcal{U}} % Ansatz
\newcommand{\pqc}{\ansatz(\params)} % PQC
\newcommand{\objective}{C} % cost operator
\newcommand{\objectiveparams}{\objective(\params)} % cost operator

\newcommand{\costfunction}{\mathcal{C}} % PQC
\newcommand{\costoperator}{\mathcal{O}} % cost operator

\newcommand{\taskidx}{\tau} % cost operator
\newcommand{\taskidxtest}{\tau^{\prime}} % cost operator
\newcommand{\stateinit}{|\psi_0 \rangle} % cost operator
\newcommand{\statefinal}{|\psi(\params) \rangle}
\newcommand{\state}{|\psi \rangle}

\newcommand{\proj}[1]{|#1\rangle \langle #1|}

\newcommand{\be}{\begin{equation}}
\newcommand{\ee}{\end{equation}}

\begin{document}

\title{FLIP: A flexible initializer for arbitrarily-sized parametrized quantum circuits}

\author{Frederic Sauvage}
\affiliation{Zapata Computing Canada Inc., 325 Front St W, Toronto, ON, M5V 2Y1}
\affiliation{Physics Department, Blackett Laboratory, Imperial College London, Prince Consort Road, SW7 2BW, United Kingdom}

\author{Sukin Sim}
\affiliation{Zapata Computing, Inc., 100 Federal Street,
Boston, MA 02110, USA}
\affiliation{Department of Chemistry and Chemical Biology, Harvard University, 12 Oxford Street, Cambridge, MA 02138, USA}

\author{Alexander A. Kunitsa}
\affiliation{Zapata Computing, Inc., 100 Federal Street,
Boston, MA 02110, USA}

\author{William A. Simon}
\affiliation{Zapata Computing, Inc., 100 Federal Street,
Boston, MA 02110, USA}

\author{Marta Mauri}
\affiliation{Zapata Computing Canada Inc., 325 Front St W, Toronto, ON, M5V 2Y1}

\author{Alejandro Perdomo-Ortiz}
\email{alejandro@zapatacomputing.com}
\affiliation{Zapata Computing Canada Inc., 325 Front St W, Toronto, ON, M5V 2Y1}

\date{\today} 

\begin{abstract}
When compared to fault-tolerant quantum computational strategies, variational quantum algorithms stand as one of the candidates with the potential of achieving quantum advantage for real-world applications in the near term. However, the optimization of the circuit parameters remains arduous and is impeded by many obstacles such as the presence of barren plateaus, many local minima in the optimization landscape, and limited quantum resources. A non-random initialization of the parameters seems to be key to the success of the parametrized quantum circuits (PQC) training. 
Drawing and extending ideas from the field of meta-learning, we address this parameter initialization task with the help of machine learning and propose FLIP: a FLexible Initializer for arbitrarily-sized Parametrized quantum circuits. FLIP can be applied to any family of PQCs, and instead of relying on a generic set of initial parameters, it is tailored to learn the structure of successful parameters from a family of related problems which are used as the training set. The flexibility advocated to FLIP hinges in the possibility of predicting the initialization of parameters in quantum circuits with a larger number of parameters from those used in the training phase. This is a critical feature lacking in other meta-learning parameter initializing strategies proposed to date. We illustrate the advantage of using FLIP in three scenarios: a family of problems with proven barren plateaus, PQC training to solve max-cut problem instances, and PQC training for finding the ground state energies of 1D Fermi-Hubbard models.
\end{abstract}

\maketitle

\section{Introduction}
\label{s:intro}

Variational quantum algorithms (VQAs) are a class of algorithms well-suited for near-term quantum computers \cite{cerezo2020variational}.
Their applications include quantum simulation and combinatorial optimization, as well as tasks in machine learning such as data classification, compression, and generation \cite{bharti2021noisy}. At the core of these near-term quantum algorithms, we encounter a parametrized quantum circuit (PQC) which acts as the quantum model we need to train to successfully perform the specific problem at hand \cite{benedetti2019parameterized}.
Optimizing PQCs remains an arduous task, and to date only optimizations over small circuit sizes have been realized experimentally. 
Several obstacles limit the scaling of VQAs to larger problems.
In particular, the presence of many local minima and barren plateaus in the optimization landscape preclude successful optimizations even for moderately small problems (see, e.g., \cite{McClean2018, cerezo2020cost,PhysRevX.10.021067}).
Furthermore, contrary to classical machine learning pipelines, the effort to obtain gradients scales linearly with the number of parameters~\cite{PhysRevA.99.032331} thus limiting the number of iterations one can realistically perform in practice.

Developing more efficient strategies for training PQCs is needed to unlock the full potential offered by VQAs and is an active topic of research \cite{Grant2019initialization,Grimsley2019,verdon2019learning,wilson2019optimizing,skolik2020layerwise}.
Drawing and extending ideas from the field of meta-learning, in particular \cite{finn2017model}, we propose to address this problem from an initialization perspective and introduce FLIP: a
\textbf{FL}exible \textbf{I}nitializer for arbitrarily-sized \textbf{P}arametrized quantum circuits. FLIP consists in a classical machine learning model that is trained on a set of PQC problems to learn how to initialize their circuit parameters in order to accelerate their optimizations.
After training, FLIP can be re-used as a boosting tool to initialize circuits corresponding to \emph{similar but new} problem instances.

The flexibility to which the name of the framework alludes takes several forms.
First, rather than relying on a generic set of initial parameters as e.g., in \cite{Grant2019initialization}, the initial parameters produced by FLIP are specially tailored for families of PQC problems and can even be conditioned on specific details of the individual problems. Secondly, the strategy is agnostic to the structure of the PQCs employed and can be used for any families of PQCs. Lastly, FLIP can accommodate circuits of different sizes (i.e., in terms of the number of qubits, circuit depth, and number of variational parameters), within the targeted family, both during its training and subsequent applications.
This is in sharp contrast with previous meta-learning parameter initialization approaches \cite{verdon2019learning,cervera2020meta}. 

We demonstrate several examples in which \zinit\ provides practical advantages.
During training, smaller circuits can be included in the dataset to help mitigating the difficulties arising in the optimization of larger ones.    
Once trained, it shows dramatically improved performances against random initialization, and also compared to a selected set of other meta-learning approaches while additionally being easier to train. 
Moreover, it is successfully applied to the initialization of larger circuits than the ones used for its training. 
This last feature is of particular appeal as \zinit\ could be trained on problem instances numerically simulated, and subsequently be used on larger problems run on a quantum device. This would allow to make the most out of cheap computational resources and to leverage the latest progress in the numerical simulation of quantum circuits in order to scale VQAs to numerically intractable problems where ultimately the advantage of VQAs is expected.

\begin{figure*}[t!]
	\includegraphics[width=0.99\textwidth]{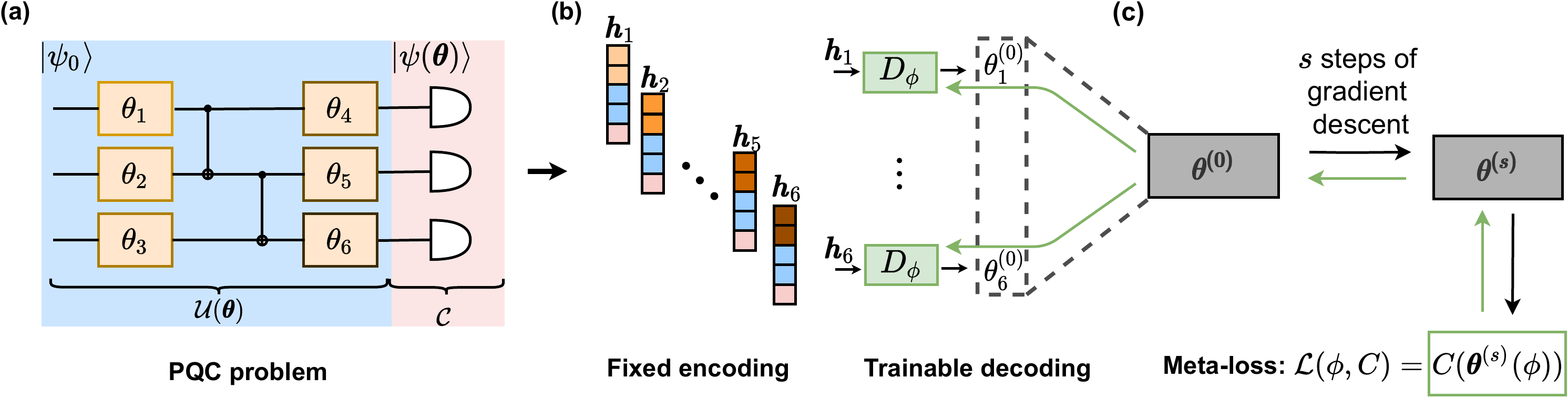}
	\caption{
    \textbf{Overview of \zinit.}
    A generic parametrized quantum circuit (PQC) problem is composed of a parametrized circuit ansatz $\pqc$ and an objective $\mathcal{C}$ which can be estimated through repeated measurements on the output state $\statefinal=\pqc \stateinit$. 
    Solving a PQC problem corresponds to the minimization of the cost function $C(\params)=\mathcal{C}(\statefinal)$.
	\textbf{(a)} Example of a PQC problem for a system size of $n=3$ qubits, and a quantum circuit with $K=6$ parameters.   
	\textbf{(b)} At the core of FLIP lies an encoding--decoding scheme which maps a PQC problem to a set of initial parameters $\init$. 
	Each of the $K$ parameters of the circuit $\mathcal{U}$ is first represented as an encoding vector $\encoding$. 
	This encoding contains information about the parameter itself (orange squares), the overall circuit (blue squares) and optionally the objective (light-red squares).
	Importantly, each of the encodings is of fixed size (here $S=5$) and uniquely represents each parameter.
	These $K$ encodings are then decoded by a neural network, $D_{\phi}$ with weights $\phi$, outputting a single value $\init_k$ per encoding $\encoding$.
	Taken together, the encoding and decoding schemes produce  a vector of initial parameters $\init$ dependent on the details of the circuits and objectives, and with dimension always matching the number of circuit parameters.
	\textbf{(c)} These parameters $\init$ are used as the starting point of a gradient descent (GD) optimization. 
	During training of \zinit, the weights $\phi$ of the decoder are tuned to output \emph{good initial parameters}, that is to minimize the meta-loss function $\mathcal{L(\phi)}$ corresponding to the value of the cost \emph{after s steps of GD}. 
	Gradients of this loss can be back-propagated (green arrows) to the weights $\phi$ of the decoder (details in the main text), which are updated accordingly. 
	In practice \zinit\ is trained over PQC problems sampled from a distribution of problems $C_\taskidx \sim p(C)$, and tested over new problems drawn from the same or a similar distribution with, for example, problems involving larger system sizes and deeper circuits.}
	\label{fig:zinit}
\end{figure*}

This work is organized as follows. In Sec.~\ref{s:flip} we describe the theoretical and practical components of FLIP. In Sec.~\ref{s:results} we discuss the main results obtained in three different scenarios where we observe a significant advantage of using our initialization scheme. In Sec.~\ref{s:outlook} we close with an outlook for potential extensions of our work.

\section{FLIP}
\label{s:flip}

The aim of \zinit\ is to accelerate optimization over targeted families of PQC problems. 
More efficient optimization is approached here from an \emph{initialization} perspective: one aims at learning a set of initial parameters which can be \emph{efficiently} refined by gradient-descent.
This point-of-view has recently emerged in the field of meta-learning \cite{finn2017model} showing promising results and has been followed by many extensions \cite{li2017meta,nichol2018first,rusu2018meta}.
However, in all these works the number of parameters to be optimized is fixed, thus precluding their applicability to circuits of different sizes.
To overcome this limitation, we introduce FLIP as a novel scheme to learn over arbitrarily-sized circuits.

We first formalize the notion of family of PQC problems in Sec.~\ref{sec:PQCproblems}, then present the technical details of FLIP.
The meta-learning aspect of the framework is reviewed in Sec.~\ref{sec:init_metalearning}, followed in Sec.~\ref{sec:encoding_decoding} by a description of the encoding--decoding scheme, i.e., the strategy we developed to be able to support circuits of arbitrary sizes.  As we will see, such scheme allows to incorporate any relevant information about the problems to be optimized, thus producing fully problem-dependent initial parameters. 

\subsection{Learning over a family of related PQC problems}
\label{sec:PQCproblems}
A generic PQC problem corresponds to a cost $\objectiveparams = \costfunction(\pqc \stateinit)$ to be minimized, where $\pqc$ denotes a parametrized circuit, with parameters $\params$, applied to an initial state $\stateinit$, and $\costfunction$ denotes an objective evaluated on the output of the circuit (Fig.~\ref{fig:zinit}(a)).
The objective is any function which can be estimated based on measurement outcomes. 
For instance, it could be the expectation value $\costfunction(\state)= \langle \psi | \costoperator | \psi \rangle$ of a Hermitian operator $\costoperator$ as it is often the case in VQAs \cite{Wecker_PRA2015,McLean2016} or a distance to a target probability distribution \cite{Benedetti2019,liu2018differentiable}. We emphasize that we have introduced a subtle distinction between the \emph{objective} $\costfunction(\state)$ which is agnostic to the circuit employed and the \emph{(PQC problem) cost} $\objectiveparams$ which is a function of the parameters $\params$ and depends both on the objective and on the choice of parametrized circuit.

Rather than considering any such PQC problem $\objective$ independently, we are interested in families of similar problems indexed by $\taskidx$ and drawn from a probability distribution, i.e., $C_\taskidx \sim p(C)$.
Such distribution can be obtained, for instance, by fixing the circuit ansatz $\ansatz$ but varying the objective $\costfunction_{\tau} \sim p(\costfunction)$,
or by fixing the objective and allowing for different circuits $\ansatz_\taskidx \sim p(\ansatz)$. 
More generally both the underlying objective and circuits can be varied.

As we will aim at exploiting meaningful parameters patterns over distributions of PQC problems (this is discussed in more details in Sec.~\ref{sec:init_metalearning} and ~\ref{sec:encoding_decoding}), we will impose some restrictions in the way these distributions are defined. 
In the following, we will consider distributions over circuits of various sizes but with the same underlying structure, and over objectives with the same attributes. 
The exact details of the distributions used are made explicit when showcasing the results.

\subsection{Initialization-based metalearning}
\label{sec:init_metalearning}
Meta-learning, i.e., learning how to efficiently optimize related problems, has a rich history in machine-learning \cite{Lemke2015,2020arXiv200405439H}. 
Here we focus on a subset of such techniques, dubbed \emph{initialization-based} meta-learners, in which the entire knowledge about a distribution of problems $p(C)$ is summarized into a single set of parameters $\init$ which is used, for any problem $C_\taskidx \sim p(C)$, as initial parameters of a gradient-based optimization.

In the original version of the method \cite{finn2017model}, these initial parameters are trained to minimize the (meta-)loss function
\be \label{eq:metaloss}
\mathcal{L}(\init) = \int \, p(C) \, C_{\taskidx}(\final)\, dC
\ee
where the parameters $\final$, for the problem $C_\taskidx$, are obtained after $s$ steps of gradient descent 
For instance, for a single step, $s=1$, of gradient descent $\pmb{\theta}_\taskidx^{(1)} = \init - \eta \nabla_{\params} C_\taskidx(\init)$. 
In practice, this number of steps is taken to be small ($s<10$) but non null. 
The case $s=0$ corresponds to finding good parameters \emph{on average} rather than good \emph{initial} parameters. 
It was shown~\cite{finn2017model, nichol2018first} that in some cases even $s=1$ can produce drastically different and better parameters than the $s=0$.

Training these initial parameters $\init$ is performed via gradient descent of the loss function Eq.~\ref{eq:metaloss}, which requires the evaluation of the gradients $\nabla_{\init} C_\taskidx(\final)$ 
where the parameters $\final$ are effectively a function of the initial parameters $\init$. These gradients can be obtained using the chain rule~\cite{nichol2018first} but involve second-order derivatives (Hessian) of the type $\nabla_{\params_i, \params_j}C_\taskidx(\params)$.
Evaluating these second-order terms is costly in general \cite{finn2017model} and even more in the context of quantum circuits \cite{PhysRevA.103.012405}. Fortunately, approximations of the gradients of the loss involving only first-order terms have been found to work well empirically, and we follow the approximation suggested in \cite{nichol2018first}
\be \label{eq:approxgrad}
\nabla_{\init} C(\final) \approx \frac{\final-\init}{\eta},
\ee 
which has been shown to be competitive and allows for a straightforward implementation.

\subsection{Encoding--decoding of the initial parameters}
\label{sec:encoding_decoding}
The meta-learning approach presented in the previous section requires the set of initial parameters $\init$ to be shared by any of the problems $C_\taskidx \sim p(C)$, which imposes the problems to have the same number of parameters, that is for PQC problems restricts its application to fixed circuits. 
However, one would expect that good initial parameters for a given PQC problem should be informative for related problems, even if of different sizes. For instance, the ground state preparation of an $N$-particle Hamiltonian probably could share some resemblance with the preparation of the ground state of a similar but extended $N+\Delta N$-particle system. 
Likewise, optimal parameters for a circuit of depth $d$ may inform us about an adequate range of parameter values for a deeper circuit of depth $d+ \Delta d$. The existence of such circuit parameters patterns, both as a function of the size of the system and of the depth of the circuit, has been observed in the context of QAOA for max-cut problems~\cite{PhysRevX.10.021067} and for the long range Ising model \cite{Pagano25396}.
This motivates us to extend the idea of learning good initial parameters for fixed-size circuits to learning good \emph{patterns} of initial parameters over circuits of varied sizes.

For this purpose we introduce a novel encoding-- decoding scheme mapping the description of a PQC problem to a vector of parameter values with the adequate dimension. The encoding part of this map is fixed, while the decoding part can be trained to produce good initial parameter values in the spirit of Sec.~\ref{sec:init_metalearning}.
Crucially, this mapping allows to condition these initial parameters upon the relevant details of the problem (circuit sizes and objectives) which are incorporated in the description of the PQC problem produced by the encoding strategy.
The general idea for a single PQC problem is illustrated in Fig.~\ref{fig:zinit}(b) and detailed in the following.

Each parameter, indexed by $k$, of an ansatz $\ansatz_\taskidx$, is \emph{encoded} as a vector $\encodingtask$ containing information about the specific nature of the parameter and of the ansatz. 
It includes, for example, the position (layers and qubits indices) and type of the corresponding parametrized gate, and the dimension (number of qubits and layers) of the ansatz. 
Several choices could be made but importantly we ensure that this encoding scheme results in encoding vectors of the same dimension $S$ for each parameter, i.e., $\forall k,\taskidx, \, dim(\encodingtask)=S$ and that distinct parameters and circuits have distinct representations, i.e., $\forall k \neq k', \, \encodingtask \neq \encodingbistask$ and $\forall \ansatz_\taskidx \neq  \ansatz_{\taskidx^\prime}, \, \encodingtask \neq \encodingtaskbis$. 
Explicit definitions of the encodings used for the results presented in Sec.~\ref{s:results} are provided in Appendix.~\ref{sec:hp:encdec}.

Once this choice of encoding is taken, any PQC problem containing an arbitrary number of parameters $K$ is mapped to $K$ of such encodings. 
These are then fed to a \emph{decoder}, denoted $D_{\phi}$, with weights $\phi$, which is the trainable part of the scheme. This decoder is taken to be a neural network with input dimension $S$ and output dimension one; that is, for any given encoding $\encodingtask$ it outputs a scalar value, and when applied to $K$ of such encodings it outputs a vector of dimension $K$ which contains the initial parameters $\inittask$ for the problem $C_\taskidx$ to be used in the meta-learning framework of Sec.~\ref{sec:init_metalearning}. Note the extra index $\taskidx$ when denoting these initial parameters $\inittask$ as they now depend on the underlying problems.

Combined together, this encoding and decoding account for a map, between a PQC problem and its initial parameters, which can be trained to learn good initial parameters as a function of the specificities of the circuits.
Additionally one can also extend the encoding to incorporate relevant information about the 	objective $\costfunction$.
This straightforward extension allows to produce fully problem-dependent initial parameters.  Finally, in Sec.~\ref{ss:training_testing} we discuss practical aspects of the training and testing of \zinit.

\subsection{Training and testing}
\label{ss:training_testing}

Training \zinit\ consists of learning the weights $\phi$ of the decoder to minimize the loss-function
\be
\mathcal{L}(\phi) = \int \, p(C) \, C_\taskidx(\final(\phi)) \, dC
\ee
which is similar to Eq.~\ref{eq:metaloss}, with the difference that the parameters $\final(\phi)$ are now obtained after $s$ steps of gradient-descent performed from the initial parameters $\inittask(\phi)$ outputted by the decoder (the dependence to the decoder weights $\phi$ has been made explicit here). 
As illustrated in Fig.~\ref{fig:zinit}(c), the gradients needed to minimize this loss function are obtained by virtue of the chain rule:
\be
\label{eq:grad_full}
\nabla_{\phi} C_{\taskidx}(\final) = \nabla_{\inittask} C_{\taskidx}(\final).\nabla_{\phi} \inittask,
\ee
where the new Jacobian term $\nabla_{\phi} \inittask$ contains derivatives of the output of the neural network $D_{\phi}$ for the different encodings $\encodingtask$, while the term $\nabla_{\inittask} C(\final)$ was discussed earlier and is approximated according to Eq.~\ref{eq:approxgrad}.
Training such hybrid schemes, involving backpropagation through quantum circuits and neural networks, has been facilitated by the recent development of several libraries \cite{bergholm2018pennylane,broughton2020tensorflow}. 
In practice each step of training of \zinit\ consists of drawing a small batch of problems from the problem distribution $p(C)$ and using the gradients in Eq.~\ref{eq:grad_full} averaged over these problems to update the weights $\phi$. 

Finally, once trained, the framework is applied to unseen testing problems. 
Testing problems, indexed by $\taskidxtest$, are sampled from a distribution $C_{\taskidxtest}\sim p'(C)$.
When presented to a new problem $C_{\taskidxtest}$ the encoding--decoding scheme is used to initialize the corresponding circuit $\ansatz_{\taskidxtest}$, from which $s'$ steps (typically larger than the number of steps $s$ used for training) of gradient descents are performed. 
In the result section, Sec.~\ref{s:results}, we draw these testing problems from distributions similar in nature to the training distribution but also containing problems of larger sizes, involving in some cases circuits twice as deep and as wide as the training ones.

\section{Results and Discussion}
\label{s:results}

\begin{figure*}
	\includegraphics[width=0.99\textwidth]{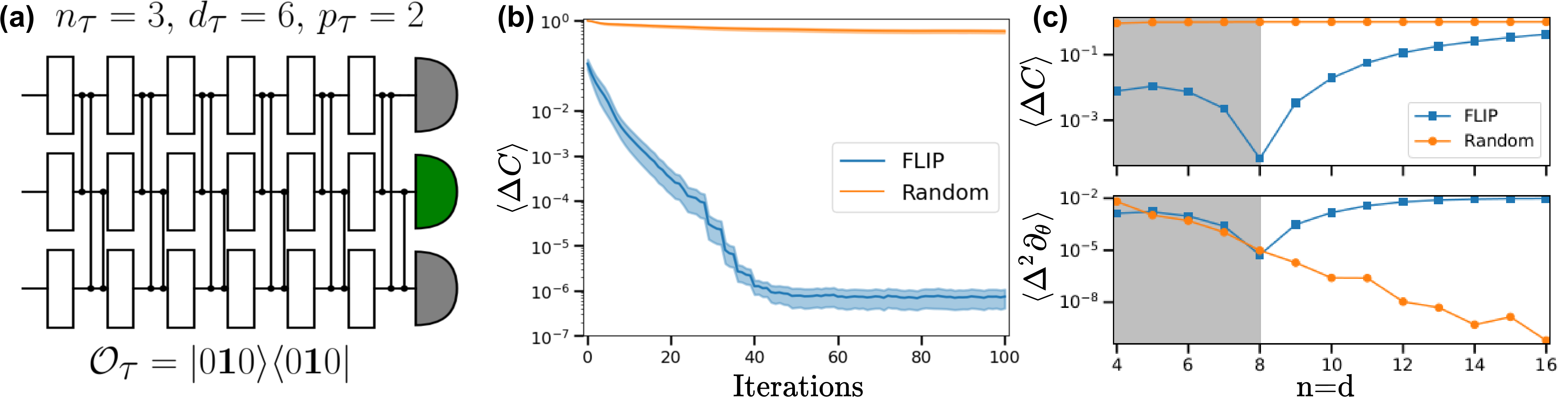}
	\caption{\textbf{State preparation problems.}
	Each problem consists in the preparation of a target state $|\psi_\taskidx^{tgt}\rangle$ of size $n_\taskidx$ with all qubits in the $|0\rangle$ state except one, at target position $p_\taskidx$, which is in the $|1\rangle$ state. 
	The objective to minimize is the negated fidelity, i.e., the expectation value of the operator $\mathcal{O_{\taskidx}}=-\proj{\psi_\taskidx^{tgt}}$. 
	Circuits are composed of $d_\taskidx$ layers of parametrized $R_y$ gates followed by fixed controlled-Z gates.
	\textbf{(a)} Problem instance for $n_\taskidx=3$ qubits, $d_\taskidx=6$ layers and target position $p_\taskidx=2$, that corresponds to the target state $|\psi_\taskidx^{tgt}\rangle=|010\rangle$.   
	\zinit\ is trained over circuits with sizes up to $n_{\taskidx}=d_{\taskidx}=8$ but tested up to the case $n_{\taskidx}=d_{\taskidx}=16$. 
	\textbf{(b)} Results of optimizations over $50$ testing problems for circuits either initialized randomly (orange curves) or with \zinit (blue curves). 
	The deviations of the objective from its minimum are reported as a function of the number of optimization steps (iterations). 
	\textbf{(c)} Initial values of the deviations in the objective (top panel) and variances in its gradients (bottom panel).  
	These are obtained for various system sizes $n$, a number of layers $d=n$ and a target position fixed to $p=1$.
	The shaded grey regions highlight system sizes used when training \zinit. 
	In the case of random initialization, the exponential decrease of the variances as a function of the system size indicates the presence of barren plateaus. 
	}
	\label{fig:barren_pres}
\end{figure*}

To put \zinit\ into practice we start with state preparation problems using simple quantum circuits, described in Sec.~\ref{s:illustr}.
This will allow us to illustrate the working details of \zinit, and the construction of a distribution of problems where both the target states and sizes of the circuits are varied.   
Furthermore, these tasks exhibit barren plateaus and we show how such issues, arising from random initialization of the circuits, could be circumvented using \zinit.

\zinit\ is then applied to some of the most promising types of VQAs encountered in the literature.
In Sec.~\ref{s:qaoa} we consider the case of the quantum approximate optimization algorithm (QAOA) \cite{Farhi2014} in the context of max-cut problems. 
As applications of QAOA to graph problems have been extensively studied \cite{brandao2018fixed,crooks2018performance,PhysRevX.10.021067,Willsch2020}, it will allow us to thoroughly benchmark \zinit\ against competitive initialization alternatives and to further investigate the patterns in the initial parameters that are learned.

In contrast to QAOA ans\"atze, hardware-efficient ans\"atze \cite{Kandala2017,Dallaire_Demers_2019} are tailored to exploit the physical connections of quantum hardware. 
Typically, they aim at reducing the depth of circuits, and thus the coherence-time requirements, at the expense of introducing many more parameters to be optimized over.
As such, if successfully optimized, they could offer practical applications in the near-term.  
In Sec.~\ref{s:fhm}, we apply \zinit\ to the ground state preparation (VQE \cite{Peruzzo2014} calculations) of the one-dimensional Fermi-Hubbard model (FHM) employing the low-depth circuit ansatz (LDCA) \cite{Dallaire_Demers_2019}. 
This FHM is treated in the half-filling regime over a continuous range of interaction strengths. 

In all these examples we demonstrate the advantage of \zinit\ compared to random initialization or other more sophisticated alternatives, and systematically assess its ability to successfully initialize larger circuits than the ones it was exposed to during training. 
All the results are obtained on numerical experiments which were carried out with Orquestra$^\text{\textregistered}$ \footnote{https://www.orquestra.io/}, Zapata's proprietary platform for workflow and data management. We also leverage here its integration with Tensorflow Quantum~\cite{broughton2020tensorflow}.

\subsection{Mitigating barren plateaus in state preparation problems}
\label{s:illustr}

Rather than considering the preparation of a single target state $|\psi^{tgt}\rangle$, we consider a family of target states $|\psi_\taskidx^{tgt}\rangle$ which are computational basis states with only one qubit in the $|1\rangle$ state, at target position $p_\taskidx$. 
This allows us to generate problems, indexed as usual by $\taskidx$, where both the size of the target state $n_\taskidx$ and the position $p_\taskidx$ can be varied. For example, for $n_\taskidx=3$ qubits, and a position $p_\taskidx = 2$, the target state reads $|\psi^{tgt}_\taskidx\rangle = |010\rangle$.

The circuits are composed of $d_\taskidx$ layers of parametrized single qubit gates $R_y(\theta)$ applied to each qubits, followed by controlled-Z gates acting on adjacent qubits (where the first and last qubits are assumed to be adjacent). The resulting parametrized circuits contain $K_{\taskidx}=n_\taskidx d_\taskidx$ variational parameters.
The objective to be minimized is taken to be the negated fidelity $C_\taskidx(\params)=\langle \psi(\params) | \costoperator_\taskidx | \psi(\params) \rangle$, with $\mathcal{O}_\tau = - |\psi_\taskidx^{tgt}\rangle\langle\psi_\taskidx^{tgt}|$. 
Distributions of problems are thus fully specified by defining how to sample the integers $n_\taskidx$, $d_\taskidx$ and $p_\taskidx$. 
A single problem with $n_\taskidx=3$ qubits, $d_\taskidx=6$ layers, and position $p_\taskidx=2$ is illustrated in Fig.~\ref{fig:barren_pres}(a).

For training, we consider a distribution of problems where the integers $n_\taskidx\in[1,8]$ qubits, $d_\taskidx\in[1,8]$ layers, and $p_\taskidx\in[1,n_\taskidx]$ are uniformly sampled within their respective range.
Further details about the hyper-parameters used during training can be found in Appendix.~\ref{sec:hp}.
For testing, $50$ new problems are sampled with $n_{\taskidxtest} \in [4,16]$ qubits, $d_{\taskidxtest} \in[4,16]$ layers, and $p_{\taskidxtest} \in[1,n_{\taskidxtest}]$, 
that is from a distribution containing problems supported by the training distribution but also larger problems (with circuits up to twice as wide and as deep as the largest circuit in the training set).

Convergence of the optimizations performed over these testing problems are depicted in Fig.~\ref{fig:barren_pres}(b) with a comparison of circuits initialized with \zinit (blue curves) and randomly initialized (orange curve). 
In both cases, $100$ steps of simple gradient descent are performed after initialization.
The absolute minimum of the objective which can be reached for these state preparation problems is $C_{min}=-1$, and we report the average (and confidence interval) of the deviation $\Delta C= C-C_{min}$ from this minimum as a function of the number of optimization steps.

One can see that circuits initialized by \zinit\ can be quickly refined to reach an average value of $\Delta C \approx 0.1\% $ after fewer than $30$ iterations. 
This is in contrast with optimizations starting with random initial parameters which even after $100$ iterations only achieve an average $\Delta C \approx 50\%$. 
These average results are dissected in Appendix.~\ref{sec:appendix:sgd} where optimization traces on individual problems are displayed (Fig.~\ref{fig:barren_sgd}). 
These individual results show that the benefit of \zinit\ is particularly appreciable for the largest circuits considered: for problems with $n_{\taskidxtest}= d_{\taskidxtest} \geq 12$, most of the optimizations starting with random parameters fail in even slightly improving the objective, while optimizations of circuits initialized with \zinit\ converge quickly.

These patterns in optimizations with randomly initialized parameters are symptomatic of barren plateaus.
To further understand the advantage of \zinit\ in this context, in Fig.~\ref{fig:barren_pres}(c) we compare the \emph{initial} values of the objective and gradients for PQCs initialized randomly (orange curves) and with \zinit (blue curves).
In the top panel the deviations $\Delta C = C-C_{min}$ of the objective values are reported while 
the variances $\Delta^2 \partial_{\theta}$ of the cost function gradients (averaged over each parameter) are displayed in the bottom panel.   
Shaded regions indicate circuit sizes seen by \zinit\ during training. 

For random initialization, the deviations of the objective value are always close to its maximum value $1$, i.e., far away from the optimal parameters.
Furthermore one can see that the amplitude of the gradients exponentially vanishes with the system size, thus preventing successful optimizations.
Circuits initialized with \zinit (blue curves) exhibit strikingly different patterns.
For problem sizes $n\leq8$ qubits, seen during training (shaded regions), both the objective and the gradient amplitudes are small, showing that \zinit\ successfully learnt to initialize parameters close to the optimal ones.
When the size of the circuits is increased further ($n>8$), the objective values increase, indicating that circuits are initialized further away from ideal parameters.
This degradation is expected as \zinit\ has to extrapolate initial parameters patterns found for small circuits to new and larger ones.
Nonetheless, in all cases the values of the initial objective stay significantly better than the ones obtained for random initialization.
More remarkably, the amplitudes of the initial gradients remain non-vanishing for the range of circuit sizes studied, thus allowing for the fast optimization results displayed in Fig.~\ref{fig:barren_pres}(b).

Complementary results are provided in Appendix.~\ref{sec:appendix:stateprep}. 
In particular, we verify that \zinit\ remains competitive even when trained and tested with noisy gradients  (Appendix.~\ref{sec:appendix:noisy}).
Overall, these results illustrate the ability of \zinit\ to learn patterns of good initial parameters with respects to the specific objective details (here corresponding to the target state to be realized) and the circuits dimensions. For these state preparation examples, where such structure exists and is relatively simple, this leads to the ability to avoid the barren plateaus phenomenon which would have arisen from random initialization.
This motivates us to apply \zinit\ further to more practical VQAs.

\subsection{Max-cut graph problems with QAOA}
\label{s:qaoa}

\begin{figure*}
	\includegraphics[width=0.99\textwidth]{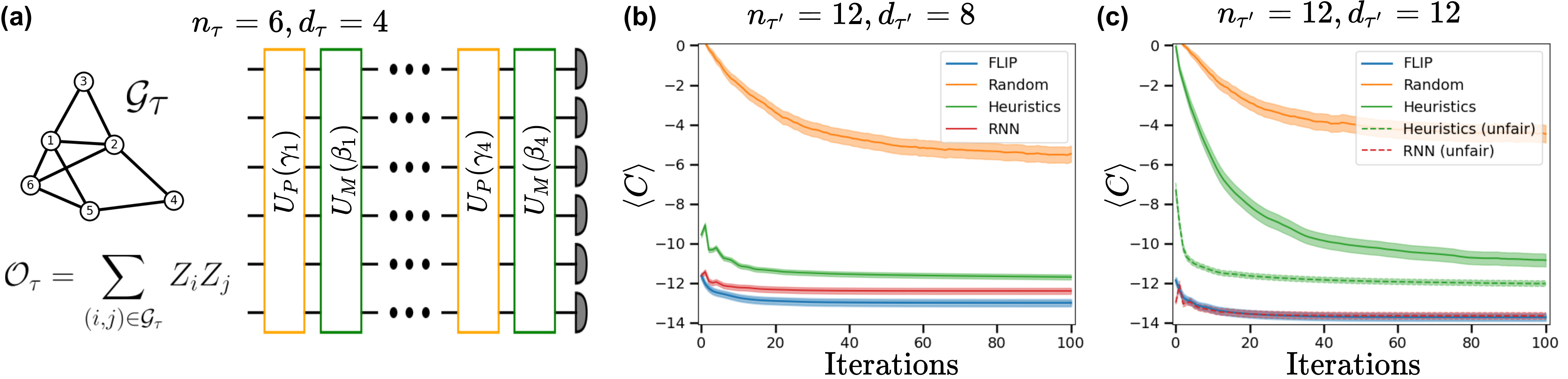}
	\caption{\textbf{
	QAOA applied to max-cut problems. }
	\textbf{(a)} The graph instances $\mathcal{G_{\taskidx}}$ are drawn from an Erdos-Renyi distribution $P_{ER}(\mathcal{G};e)$, with the parameter $e\in[30\%, 90\%]$ specifying the probability of any edge to be included. Example for a graph with $n_\taskidx= 6$ nodes and probability $e=50\%$.
	Solving the max-cut of the graph $\mathcal{G_{\taskidx}}$ can be mapped to the minimization of the expectation value $\langle \costoperator_\taskidx \rangle$ of the cost operator defined in the figure.    
	This ground state preparation is attempted with a QAOA ansatz consisting of $d_{\taskidx}$ layers, of parametrized problem (orange) and mixer (green) unitaries. 
	\textbf{(b-c)} Different initialization strategies (described in the main text) are compared: \zinit, recurrent neural network based meta-learning (RNN), heuristics initial parameters, and random initialization. 
	Except for the case of random initialization each of the initialization strategies is trained beforehand. 
	For \zinit\ this training is performed over circuits with $d_{\taskidx} \leq 8$ layers, while RNN and heuristics are trained on circuits containing $d_{\taskidx} = 8$ layers (as they cannot be trained on circuits of varying depths). 
	Optimization results averaged over $100$ new testing graph instances containing $n_{\taskidxtest}=12$ nodes and circuits with (b) $d_{\taskidxtest}=8$ (b) and (c) $d_{\taskidxtest}=12$ layers. 
	For FLIP, the latter case corresponds to testing over more parameters than initially trained for. The RNN needs to be re-trained from scratch on these larger circuits and, thus, labelled \textit{unfair} since it is given this advantage over FLIP which does not use training with these larger instances to make its predictions. 
	The heuristics initial parameters trained for $8$ layers can be padded with random values, or new heuristics parameters can be re-trained on the larger circuits (thus labelled unfair).  
	\zinit\ is not re-trained but still outperforms other initialization strategies.   
	}
	\label{fig:qaoa_pres}
\end{figure*}

The Quantum Approximate Optimization Algorithm (QAOA) \cite{Farhi2014} was suggested as an approximate technique for optimizing combinatorial problems. Since its proposal, QAOA has received a lot of attention with recent works focusing on aspects of its practical implementation and scaling \cite{PhysRevX.10.021067, PRXQuantum.1.020304, Harrigan2021, Pagano25396}.
Alternating-type ans\"atze such as QAOA as well as the Hamiltonian Variational Ansatz \cite{Wecker_PRA2015,wiersema2020exploring} have the advantage of being parameter-efficient.
Still, optimizing such ans\"atze can be challenging and it was found that even for small problem sizes, the optimization landscape is filled with local minima \cite{PhysRevX.10.021067, Willsch2020}.
This has motivated many works \cite{wilson2019optimizing,verdon2019learning,PhysRevX.10.021067,PhysRevResearch.2.023074} aiming at devising more efficient optimization strategies.
We first briefly recall the definition of max-cut problems and of the QAOA ansatz, then apply \zinit\ and compare it to random initialization and other more sophisticated initialization strategies.

Consider a graph $\mathcal{G}$ with a set of $n$ nodes $\mathcal{V}$ and a set of edges $\mathcal{E}$.
The maximum cut of this graph is defined as the partition $(\mathcal{V}_1, \mathcal{V}_2)$ of $\mathcal{V}$, which maximizes the number of edges having both an end-point in $\mathcal{V}_1$ and $\mathcal{V}_2$.
A max-cut problem can be mapped to the $n$--qubit operator $\costoperator = \sum_{(i,j)\in \mathcal{E}} Z_i Z_j$, whose ground state provides a solution to the problem.
While in principle this ground state preparation could be attempted with any type of PQCs, it is typical to resort to QAOA ans\"atze for these problems.
A QAOA ansatz is formed of repeated layers of \emph{problem} and \emph{mixer} unitaries defined respectively as $U_P(\gamma)=exp(-i \gamma \costoperator)$ and $U_M(\beta)=exp(-i \beta \sum^n_{i=1} X_i)$.
For $d$ of such layers the overall ansatz reads $\pqc=\prod^1_{l=d} U_M(\beta_l) U_P(\gamma_l)$ where $\params=(\gamma_1, \beta_1, \hdots, \gamma_d, \beta_d)$ is the set of the $K=2d$ parameters to be optimized over.

Graph instances $\mathcal{G}_{\taskidx} \sim P_{ER}(\mathcal{G};e)$ are drawn from an Erdos-Renyi distribution with parameter $e$.
This parameter $e$ specifies the probability of any edge to belong to the graph.
We follow \cite{verdon2019learning} and rather than keeping this probability fixed we also sample it uniformly from $e_\taskidx \in [30\%, 90\%]$ each time a new graph is drawn.
When training \zinit\, the number of graph nodes $n_\taskidx \in [2,8]$ and the number of circuit layers $d_\taskidx \in [2,8]$ are both uniformly sampled. 
An instance of a QAOA max-cut problem belonging to the training data set is illustrated in Fig.~\ref{fig:qaoa_pres}, for the case where $d_\tau=4$ layers, $n_\taskidx=6$ nodes and $e_\taskidx=50\%$.  

As in the previous case, we will compare optimizations with circuits initialized by \zinit\ against random initializations, but we will also include more competitive baselines.  
In \cite{brandao2018fixed}, it was reported that the objective values of QAOA ans\"atze concentrate for fixed parameters but different problem instances. 
In other words, good parameters obtained for a given graph are typically also good for other similar graphs.
While these results are obtained for $3$--regular graphs and a number of layers smaller than the number of nodes, this motivates us to build a simple general initialization strategy.
This strategy - that we call \emph{heuristics initialization} - consists in:
\begin{itemize}
    \item[(i)] performing optimization over randomly drawn training problems,
    \item[(ii)] selecting the set of optimal parameters resulting in the best \emph{average} objective value over the training problems, 
    \item[(iii)] reusing these parameters as initial parameters when optimizing new problems.
\end{itemize}
The two-step training part of the strategy allows to mitigate for (i) optimizations trapped in local minima by repeated optimizations, and (ii) to ensure that the selected parameters are typically good for other problems.

In addition, we also include results obtained with the recurrent neural network (RNN) meta-learner approach \cite{verdon2019learning}.
A RNN is trained to act as a black-box optimizer: at each step it receives the latest evaluation of the objective function and suggests a new set of parameters to try. 
After its training, this RNN can be used on new problem instances for a small number of steps. The best set of parameters found over these preliminary steps is subsequently used as initial parameters of a new optimization. For its implementation we follow \cite{verdon2019learning}.

In contrast with \zinit\, both the heuristics and the RNN initializer require that all the problem instances share the same number of parameters. 
For QAOA circuits, this restricts the circuits employed to be of fixed depth, although the number of graph nodes can be freely varied as it does not relate directly to the number of parameters involved.
Hence, when training these alternative initializers, we resort to a similar training distribution as the one used for \zinit, with the exception that all circuits are taken to be of fixed depth, $d_\taskidx =8$ layers. 

A first batch of testing problems are generated for graphs with $n_{\taskidxtest}=12$ nodes and circuits with $d_{\taskidxtest}=8$ layers.
Average optimization results (and confidence intervals) over $100$ of such testing problem instances are reported in Fig.~\ref{fig:qaoa_pres}(b).
The four different initialization strategies previously discussed are compared. 
One can see that the simple heuristic strategy already provides a significant improvement compared to random initialization, thus highlighting the importance of informed initialization of the circuit parameters.
An extra improvement is achieved when using the RNN initializer. 
Finally circuits initialized with \zinit\ exhibit the best final average performance over these problems. 
While the initial objective values are similar for circuits initialized by \zinit\ and RNN, the initial parameters produced by \zinit\ are found to be more auspicious to further optimization.

Results for new testing problems with an increased depth of $d_{\taskidxtest}=12$ layers are displayed in Fig.~\ref{fig:qaoa_pres}(c).
The heuristics initial parameters trained on circuits with $d_{\taskidx}=8$ layers, are adapted to these larger circuits (with increased number of parameters) by padding the missing parameters entries with random values.
However, there is no straightforward way to fairly extend the RNN trained on circuits with $d_{\taskidx}=8$ layers to these larger circuits. 
Hence, we include the heuristics and RNN initializer re-trained from scratch on problems with $d_{\taskidx}=12$ layers.
These are labeled as ``unfair'' in the legend as they are trained on circuits $50\%$ deeper than the largest ones seen by \zinit\ during its training.
Remarkably, even in this challenging set-up, \zinit\ outperforms all the other approaches, albeit only showing an almost indistinguishable advantage compared to the RNN trained on the $d_{\taskidx}=12$ layered circuits ($\Delta C \approx 0.06$).

We also investigate the patterns in the initial parameters found by the framework.
These are plotted for different circuit sizes ranging from $d=2$ to $d=15$ layers, in Fig.~\ref{fig:qaoa_patterns} of the Appendix.~\ref{sec:supp:qaoa}. 
In particular, we found similar patterns as the ones discovered and exploited in \cite{PhysRevX.10.021067}. In contrast to \cite{PhysRevX.10.021067}, these patterns are learnt during a single phase of training, without requiring neither sequential growing of the circuits nor the use of handcrafted extrapolation rules.

Having shown the benefits of \zinit\ for initializing QAOA circuits, we now consider its application to a circuit ansatz with a more involved structure.

\subsection{Initializing LDCA for the 1D Fermi-Hubbard Model}\label{s:fhm}
\begin{figure*}[t!]
	\includegraphics[width=0.99\textwidth]{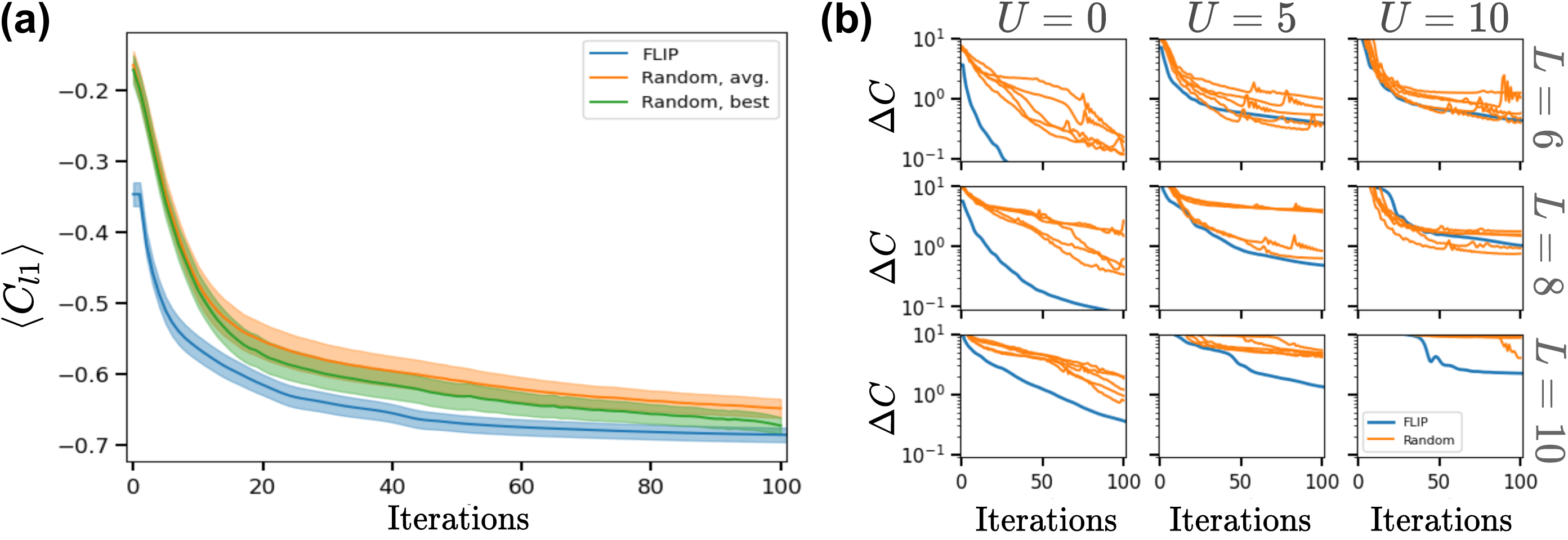}
	\caption{
    \textbf{Optimization results for the 1D Fermi-Hubbard model (1D FHM).} 
	\zinit\ is trained on LDCA ans\"atze with a number of layers $d_\taskidx \leq6$ and a number of sites  $L_\taskidx \leq 6$. All 1D FHM instances are in the  half-filling regime with positive interaction strength $U\in[0,10]$.
	For testing these numbers are increased to $d_{\taskidxtest} = 8$ and  $L_{\taskidxtest} \in \{6,8,10\}$. 
	Optimizations of the circuits initialized by \zinit (blue curves) or randomly (with $5$ restarts) are compared. 
	\textbf{(a)} Convergence of the $l1-$ normalized energies $\langle C_{l1} \rangle$, averaged over the testing problems. 
	For random initialization the average (orange curve), and the best per problem over the $5$ restarts (green curve) are reported.
	\textbf{(b)} Results for individual testing problems with interaction strengths $U=0,5$ and $10$ (from left to right column) and a number of sites $L=6,8$ and $10$ (from top to bottom row).
	Deviations $\Delta C = C - C_{min}$ of the optimized raw energies from the true ground state energies are reported. 
	Each of the five traces correspond to the repeated random initialization (orange curves).
}
	\label{fig:fhm}
\end{figure*}

The Fermi-Hubbard model (FHM) is a prototype of a many-body interacting system, widely used to study collective phenomena in condensed matter physics, most notably high-temperature superconductivity~\cite{Dagotto1994}. Despite its simplicity, FHM features a broad spectrum of coupling regimes that are challenging for the state-of-the-art classical electronic structure methods~\cite{LeBlanc2015}. In the context of VQAs, various classes of FHMs were used to benchmark VQE optimizers~\cite{wilson2019optimizing} and parameter initialization heuristics~\cite{verdon2019learning}. Recently Cade et al.~\cite{Cade2020} analyzed the prospect of achieving quantum advantage for the large scale VQE simulations of the two-dimensional FHM, emphasizing the need for efficient circuit parameter optimization techniques, including those based on meta-learning.
Here we focus on the one-dimensional FHM (1D FHM), which describes a system of fermions on a linear chain of sites with length $L$.
The 1D FHM Hamiltonian is defined as the following in the second quantization form:
\begin{equation}
\label{eq:fhm_hamiltonian}
\begin{split}
    H_{\text{1D FHM}} = &-t \sum_{\sigma=\uparrow, \downarrow} \sum_{j=1}^{L-1} 
    (a^\dag_{j+1, \sigma} a_{j, \sigma} + a^\dag_{j, \sigma} a_{j+1, \sigma} ) \\
    &+ U \sum_{j=1}^{L} n_{j, \uparrow} n_{j, \downarrow} - \mu \sum_{\sigma=\uparrow, \downarrow} \sum_{j=1}^{L} n_{j, \sigma},
\end{split}
\end{equation}
where $j$ indexes the sites and $\sigma$ indexes the spin projection. 
The first term quantifies the kinetic energy
corresponding to fermions hopping between nearest-neighboring sites and is proportional to the tunneling amplitude $t$.
The second term 
accounts for the on-site Coulomb interaction with strength $U$. 
Symbols $n_{j, \sigma}$ refer to number operators.
Lastly, the third term is the chemical potential $\mu$ that determines the number of electrons or the filling. 
For the half-filling case, in which the number of electrons $N$ is equal to $L$, $\mu$ is set to $\frac{U}{2}$. 

For the infinite 1D FHM, the ground state energy density per site is exactly solvable using the Bethe ansatz \cite{lieb2003one}.
Consequently, equipped with verifiable results, this model has been proposed as a benchmark system for near-term quantum computers \cite{dallairedemers2020application}. 
In this work, ground state energies of the 1D FHM over
a range of chain lengths are systematically estimated using the VQE algorithm. 
With increasing system size (chain length) and corresponding increase in the circuit resources (e.g., depth or gate count) of the respective VQE ansatz, the noise in the quantum device deteriorates the quality of the solutions. 
The maximum chain length before the device noise dominates the quality of VQE solutions informs the maximum capability of the particular quantum device at solving related algorithm tasks. 

Implementation of such VQE benchmark on near-term devices requires a careful design of a variational ansatz with low circuit depth.
A candidate for such ansatz is the ``Low-Depth Circuit Ansatz'' (LDCA), a linear-depth hardware-inspired ansatz for devices with linear qubit connectivity and tunable couplers \cite{Dallaire_Demers_2019}. 
While LDCA was shown to be effective in estimating ground state energies of strongly correlated fermionic systems, its application has been limited to small problem sizes due to the quadratic scaling of parameters with the system size and the corresponding difficulty in parameter optimization. 
A recent work proposed an optimization method for parameter-heavy circuits such as LDCA, but the reported simulations for LDCA required many energy evaluations on the quantum computer \cite{sim2020adaptive}.  
This prompts a need for better parameter initialization strategies to reduce the number of energy evaluations. 
For a parameter-heavy ansatz like LDCA, in which the role of each parameter (and its corresponding gate) is not easily understood,
it would especially be beneficial to have an initialization strategy that is effective across a family of related problem instances.
This poses an opportunity for a strategy like FLIP.

In the following paragraphs, we describe details and results for applying FLIP to initialize number-preserving LDCA for 1D FHM problem instances of varying chain lengths $L$ and numbers $d$ of circuit layers  (named ``sublayers'' in LDCA). 
The structure of LDCA used in this study, including the definition of a sublayer, can be found in the Appendix of Ref. \cite{sim2020adaptive}. In all cases the ansatz circuit is applied to non-interacting anti-ferromagnetic initial states with two electrons per occupied lattice site.
For this version of LDCA, which conserves the particle number, there are $K=3 d (n-1) + n$ parameters where the system size $n=2L$, i.e., two qubits are used per lattice site.

Training instances for \zinit\ were generated for a number of sites $L_{\taskidx} \in [1,6]$, a value of the interaction $U_{\taskidx} \in [0,10]$ and a number of LDCA sublayers $d_{\taskidx}\in[1,6]$. 
For each new training problem, these values are sampled uniformly within their respective discrete or continuous ranges and the cost function is taken to be to the expectation value of the $l1$-normalized version of the problem Hamiltonian, Eq.~\ref{eq:fhm_hamiltonian}.
For testing, both the number of sites and of circuit layers are increased to $L_{\taskidxtest} \in \{6, 8, 10\}$ and $d_{\taskidxtest}=8$ and values of $U$ are taken at regular intervals in the range $[0,10]$.  
Extended details on the training and testing can be found in Appendix.~\ref{sec:hp}.

Optimization results averaged over the testing problems are reported in Fig.\ref{fig:fhm}(a).
For random parameter initialization, each problem optimization is restarted five times.
Results corresponding to the average over these repetitions (orange) or to the best run per problem (green) are compared to optimizations of circuits initialized with \zinit (blue).
After only \emph{c.}$30$($50$) steps of optimization, circuits initialized with \zinit\ achieve similar convergence when compared to $100$ steps of optimizations from random initialization for the average (best of five) case. 
When looking at individual cases, as reported in Fig.\ref{fig:fhm}(b), one can see that again the advantage of \zinit\ is the most prominent for the largest circuits (last row) over which it was applied. Importantly, \zinit\ outperforms random initialization in the strong coupling regime (i.e., away from $U = 0$), where the non-interacting initial state generally provides a poor starting point for VQE optimization. 
We emphasize that prior to this study there was no method for initializing parameters of LDCA circuits other than assigning random values.

Finally, in Appendix.~\ref{sec:extra:fhm} we report results obtained for an extended number of optimization steps performed after initialization. We also include results for the case where the interaction strength can adopt both positive and negative values. 
In particular, we obtained similar positive results for the case $U \in [-3,10]$ but less of an advantage found on the extended range $U \in [-10,10]$. There, we discuss potential modifications to FLIP to boost its performance in that region as well.

\section{Outlook}\label{s:outlook}

In its simplest form, \zinit\ can be applied to a single VQA objective but with circuits of varying depths.
Rather than growing the circuits sequentially and making
incremental adjustments of parameters \cite{skolik2020layerwise, PhysRevX.10.021067, Pagano25396}, which may fail in certain cases \cite{campos2020abrupt},
\zinit\ aims at capturing and exploiting patterns in the parameter space 
and thus can provide a more robust approach. 
This flexibility towards learning over circuits of different sizes is one of the outstanding features of our initialization scheme. Still, as illustrated in the three case studies presented here, the full capability of \zinit\ appear in scenarios where both the circuits and the objectives (corresponding to the target states in Sec.~\ref{s:illustr}, the graph instances in Sec.~\ref{s:qaoa}, and the interaction strengths in Sec.~\ref{s:fhm}) are varied. 
Rather than treating each problem individually, \zinit\ provides a unified framework to learn good initial parameters over many problems, resulting in overall faster convergence.

Although the first use case of applying FLIP to mitigate barren plateaus was demonstrated in a simple synthetic setting similar to those previously studied in the literature, we further demonstrated that FLIP is a promising initialization technique for more complex problems. The intuition behind a successful parameter initialization with FLIP is that as long as there is some (hidden) structure in the parameter space that can be learned by the framework, this can be exploited to adequately initialize new circuits even when these circuits are larger than the ones seen during training. A clear demonstration of learning such patterns was the application of \zinit\  to the max-cut instances in QAOA, in which it  outperformed other proposed initialization techniques. We also observed an enhancement over random initialization in the application to the 1D FHM instances, where the structure in the parameter space is not obvious even after training. We are currently working on extending \zinit\ to other application domains, 
especially those that lack 
a ``problem Hamiltonian'' to guide the construction of the circuit ansatz. This is, for example, the case for probabilistic generative modeling with Quantum Circuit Born Machines (QCBMs)~\cite{Benedetti2019}.

We highlight that our proposed encoding--decoding scheme allows to fully condition the initial parameters with respect to specific details of the problem athand, e.g., the interaction strength of the parametrized Hamiltonians in Sec.~\ref{s:fhm}. 
Such problem-dependent feature was also proposed in \cite{cervera2020meta} but their method was limited to fixed-size circuits and relied on a more rigid encoding scheme.
This possibility to easily incorporate informative details of any problem can be further explored and exploited. For the QAOA graph problems, we intend to extend the encoding of Sec.~\ref{s:qaoa} to also incorporate information about the graph instances, e.g., their densities.

The meta-learning aspect of \zinit\ we have 
adapted here~\cite{finn2017model} is a well-studied paradigm for which many extensions have been proposed \cite{li2017meta,nichol2018first,rusu2018meta,flennerhag2019meta}.  
These could be readily incorporated. In particular, the ability to train the learning rates to be used after initialization \cite{li2017meta}, in addition to training the initial parameters, could further contribute to more efficient optimizations (this is discussed further in Appendix.~\ref{sec:extra:fhm}).   

Finally, we note that a recent work reported well-behaved optimization landscapes for over-parametrized circuits in the case of employing the Hamiltonian Variational Ansatz \cite{wiersema2020exploring}.
In this work, the authors observed that for circuits with depths scaling at most polynomially with the system size, low-quality local minima in corresponding objective landscapes disappear, and optimization becomes relatively easy. 
It would be interesting to apply \zinit\ to these problems and assess if this onset of ``easy trainability'' can be even further enhanced with better initialization strategies. 
In general, we expect that informed initialization of the parameters can accelerate convergence
and thus reduce the overall number of circuits to be run, which is critical for extending the application of VQAs to larger problem sizes. As gate-based quantum computing technologies mature, initialization techniques which embrace this unique flexibility will be essential to mitigate the challenges in trainability posed for PQC-based models and eventually scale to their application in real-world applications settings.

\begin{acknowledgments} 

F.S would like to acknowledge Zapata Computing for hosting his Quantum Applications Internship. All authors would like to acknowledge access to the Orquestra$^\text{\textregistered}$ software platform where all simulations where performed. 
\end{acknowledgments}

\appendix
\section{Hyper-parameters}
\label{sec:hp}

In this appendix we report the design choices, commonly referred as hyper-parameters, used for the results presented in Sec.~\ref{s:results}.
For both the state preparation example of Sec.~\ref{s:illustr} and the FHM of Sec.~\ref{s:fhm}, \zinit\ is only compared to random initialization.
For the max-cut problems of Sec.~\ref{s:qaoa}, \zinit\ is compared to other trainable initialization strategies (heuristics and RNN) for which the hyper-parameters used are also reported here.
These hyper-parameters are grouped in three categories: Encoder-Decoder for \zinit (\ref{sec:hp:encdec}), training of the initializers (\ref{sec:hp:train}) and testing (\ref{sec:hp:test}). 

\subsection{Encoder -- Decoder}
\label{sec:hp:encdec}
As detailed in Sec.~\ref{sec:encoding_decoding}, the encoding part of \zinit\ consists of a fixed mapping from any parameter belonging to a PQC problem to an encoding vector of size $S$.
Crucially, this encoding vector is taken such that any pair of parameter and circuit is mapped to a distinct vector.
These encodings are then fed to a trainable decoder producing initial values of the parameters.

For the state preparation problems in Sec.~\ref{s:illustr}, each encoding vector contains the location of the parametrized gate, i.e., its qubit and layer index, and the size of the overall ansatz, i.e., its number of qubits and layers. In addition to these $4$ values, the target position which specifies the state to be prepared $p_\tau$ is also included, accounting for a total size of $S=5$ values per encoding.   

For the QAOA circuits in Sec.~\ref{s:qaoa}, the layer index, the total number of layers, and a Boolean value indicating the gate type (problem or mixer unitary) are part of the encoding, giving a total of $S=3$ entries.

At last, for the LDCA circuits in Sec.~\ref{s:fhm}, each encoding includes the index of the first qubit the corresponding parametrized gate acts on, the type of gate, and its layer index. In this case, eight different types of gates are involved and are encoded using a one-hot encoding of size $8$. 
The eight gate types are the following:
\begin{itemize}
    \item $R_z$ acting on initial states (1) $|0 \rangle$ and (2) $|1 \rangle$,
    \item Gate operations $e^{-i \theta XX/2}$ or $e^{-i \theta YY/2}$ in (3) even and (4) odd circuit layers \footnote{We refer to a layer of nearest-neighboring two-qubit gates that start with acting on the first and second qubits as an even layer, and one that start with acting on the second and third qubits as an odd layer.},
    \item Gate operation $e^{-i \theta ZZ/2}$ in (5) even and (6) odd layers,
    \item Gate operations $e^{-i \theta XY/2}$ or $e^{-i \theta YX/2}$ in (7) even and (8) odd layers.
\end{itemize}
We refer the readers to \cite{sim2020adaptive} for further details on the construction of the ansatz.
Additionally, the size of the ansatz, i.e., its number of layers and qubits, and the value of the interaction strength $U$ are also included, giving an overall dimension of $S=13$. 

The decoder consists of a simple feed-forward neural network, with ReLu activation functions and linear output, of dimensions (number of layers $\times$ neurons per layer) $6 \times 30$ for the state preparations, $4 \times 30$ for the max-cut, and $4 \times 20$ for the FHM problems.

In machine learning, it is good practice to make sure that inputs and outputs of neural networks have reasonable scale.
For this purpose, the outputs of the decoder, which correspond to rotation angles, are systematically rescaled by a factor $\pi$ and elements of the encoding vectors (except for the gate types) are divided by a factor in between $10$ to $15$, as we consider circuits up to maximum of c.$20$ qubits and layers. Finally in all the examples the costs used during training are normalized using the $l1$-norm of the corresponding operators.

\subsection{Training}
\label{sec:hp:train}
For training, we report in Table.~\ref{tab:train} the number $N$ of problem instances used and their sizes, i.e., the number of qubits $n$, the depth $d$, and the number of parameters $K$ of the circuits. 
In the case of \zinit\ these values are indicated as a range, as training is performed on problems of different sizes.
In addition we include the learning rate $\alpha$ and number of epochs $e$ used for training. All the initializers are trained using Adam \cite{kingma2014adam} with learning rate $\alpha$. 
For \zinit\ the value $s$, used in Eq.~\ref{eq:metaloss}, corresponding to the meta-learning aspect is fixed to $5$, and $\eta$ is reported in Table~\ref{tab:train}.

\begin{table}[h]
\caption{Training hyper-parameters.} 
\centering
\begin{tabular}{r ccccccc} 
\hline\hline               
 &$N$ & $n$ &$d$ &$K$ & $e$ &$\alpha$ &$\eta$ \\ [0.5ex]   
\hline
\textbf{State preparation} \quad& & & & & & &  \\
\emph{FLIP}   & 150 & [1,8]& [1,8]& [1,64]& 100&$4.10^{-3}$& $10^{-1}$ \\
\textbf{QAOA}   \,\,\,\quad\quad\quad\quad\quad& & & & & & &  \\
\emph{FLIP}   & 200 & [6,9]& [1,8]& [1,16]& 90 &$4.10^{-3}$& $10^{-1}$ \\
\emph{Heuristics}   & 200 & [6,9]& 8& 16& &$4.10^{-3}$& $10^{-1}$ \\
\emph{RNN}   & 200 & [6,9]& 8& 16& 400&$4.10^{-3}$& $10^{-1}$ \\
\textbf{Fermi-Hubbard}   \,\,\,\quad& & & & & &  \\
\emph{FLIP}   & 300& [4,12]& [2,6]& [8,276]& 100&$10^{-3}$& $2.10^{-2}$ \\
\hline 
\end{tabular}\label{tab:train}
\end{table}

\subsection{Testing}
\label{sec:hp:test} 

Similarly, in Table~\ref{tab:test} we report the number of problem instances $N$ used to produce the averaged testing results and their sizes. 
After initialization (either with \zinit, randomly or other initializer presented in Sec.~\ref{s:qaoa}), optimizations are performed for $100$ steps with standard gradient descent (\ref{s:illustr}) or Adam (Sec.~\ref{s:qaoa}, \ref{s:fhm}). 
The learning rates $\alpha$ used are also reported. 
In each case we ensured that appropriate learning rates were chosen. For random initialization, these were found to be larger than the rates used after informed initialization.

\begin{table}[h]
\caption{Testing hyper-parameters.}
\centering
\begin{tabular}{r ccccc}
\hline\hline            
 &$N$ & $n$ &$d$ &$K$ &$\alpha$ \\ [0.5ex]   
\hline
\textbf{State preparation} \,\,\quad& & & & &  \\
\emph{FLIP}   & 50 & [4,16]& [4,16]& [22,210]& $10^{-1}$ \\
\emph{Random}   & \texttt{"} & \texttt{"}& \texttt{"}& \texttt{"}& $3.10^{-1}$ \\
\textbf{QAOA}   \quad\quad\quad\quad\quad\quad& & & & &   \\
\emph{FLIP}   & 100 & 12& 8(12)& 16(24) & $2.10^{-2}$ \\
\emph{Heuristics}   & \texttt{"} & \texttt{"}& \texttt{"}& \texttt{"}& $2.10^{-2}$ \\
\emph{RNN}   &  \texttt{"}& \texttt{"}& \texttt{"}& \texttt{"}&$2.10^{-2}$ \\
\emph{Random}   & \texttt{"}& \texttt{"}& \texttt{"}& \texttt{"}&$10^{-1}$ \\
\textbf{Fermi-Hubbard}   \quad\quad& & & &   \\
\emph{FLIP}   & 21& [12,20]& 8& [268,476]& $2.10^{-2}$ \\
\hline
\end{tabular}\label{tab:test}
\end{table}

\section{State preparations}
\label{sec:appendix:stateprep}
In Sec.~\ref{s:illustr} we compared optimizations for state preparation problems, ran for PQCs initialized both randomly and with \zinit.
Results presented in Fig.~\ref{fig:barren_pres} show the advantage of \zinit\ over random initialization on average over $50$ testing problems.
Here we further characterize this advantage. First, in Appendix.~\ref{sec:appendix:vanishgrads}, we present extended data exhibiting vanishing gradient patterns, i.e., barren plateaus, for randomly initialized circuits.
Then, individual optimization results are reported in Appendix.~\ref{sec:appendix:sgd}.
In line with Sec.~\ref{s:illustr}, these results are obtained with standard gradient descent routines, however we notice that one could seemingly escape barren plateaus in the case of random initialization by using more refined optimization techniques, i.e., Adam \cite{kingma2014adam}. This is reported and discussed in Appendix.~\ref{sec:appendix:adam}. 
However, as showcased in Sec.~\ref{sec:appendix:noisy}, when noise is taken into account, this workaround collapses.
Still, even with the addition of noise during training and testing, \zinit\ remains highly competitive.

All the results in this appendix section correspond to the problems presented in Sec.~\ref{s:illustr} and, except when explicitly stated otherwise, we resort to the exact same version of \zinit, that is without any additional training. 
For the sake of simplicity we only illustrate results for the case where the target position is fixed to $p=1$, which corresponds to the task of preparing a state $|10 \hdots 0\rangle$ of arbitrary size, but other values of $p$ exhibit similar performances.

\subsection{Vanishing gradients}
\label{sec:appendix:vanishgrads}

\begin{figure}
	\includegraphics[width=0.49\textwidth]{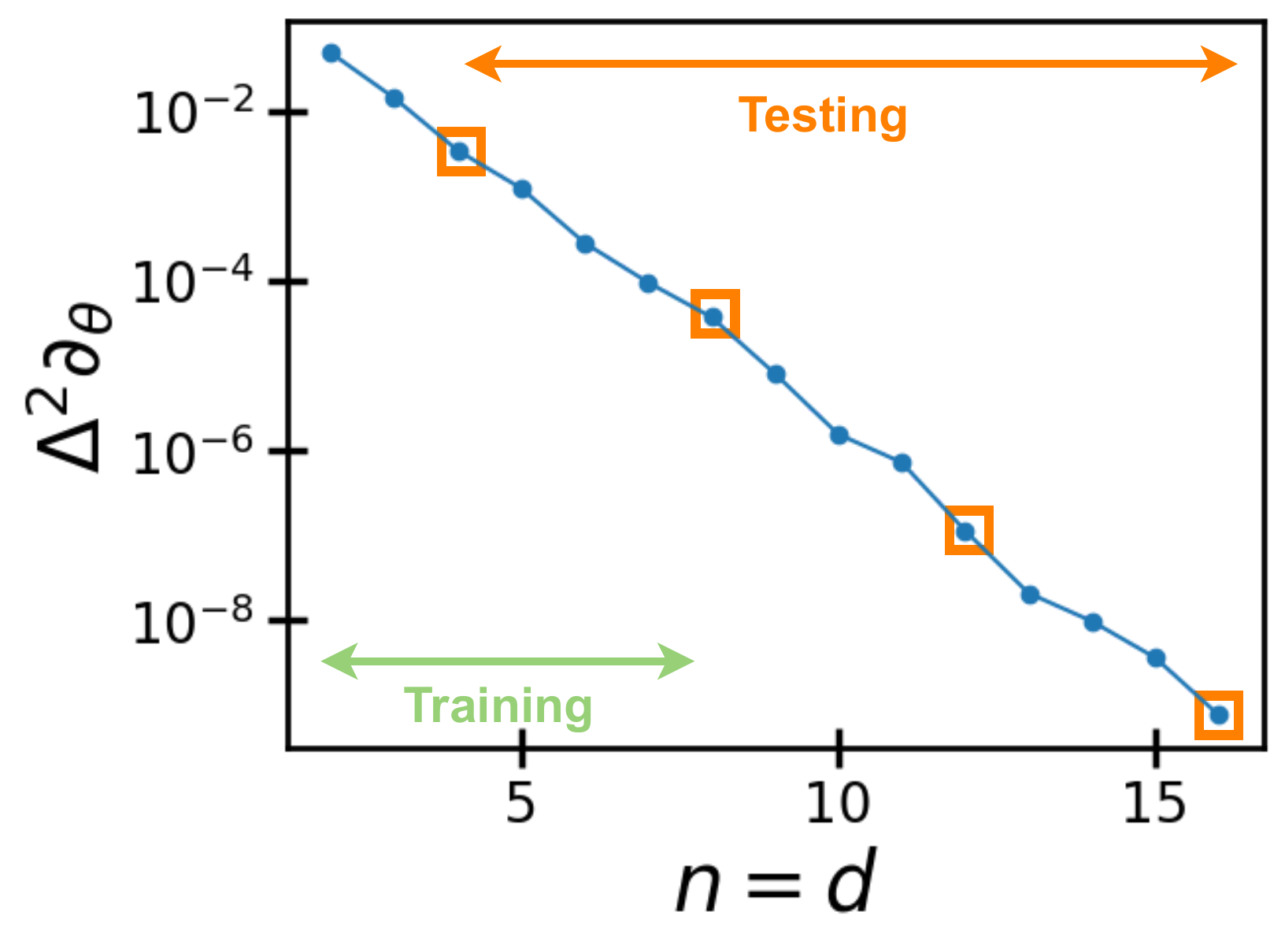}
	\caption{
	Empirical variances in the gradients of the state preparation objectives for randomly initialized circuits. 
	These variances are obtained for different system sizes from $n=2$ to $16$ qubits and a circuit depth scaling linearly with the size, $n=d$.  
	Each circuit is repeated $250$ times with random parameters, and we report the average of the variances of the gradient $\Delta^2 \partial_\theta$ over each parameter.
	Optimization of the circuits with sizes highlighted in orange squares are studied individually in Fig.~\ref{fig:barren_sgd},\ref{fig:barren_adam}, \ref{fig:barren_noise}. 
	}
	\label{fig:vanish_grads}
\end{figure}

	In Fig.~\ref{fig:vanish_grads}(c) we report empirical variances of the gradients as a function of the system size.
	For these data we take the number of layers to be equal to the system size.
	Variances $\Delta^2 \partial_\theta$ are obtained over $250$ random initializations for each circuit size and averaged over each of the parameters.
	We observe an exponential decay 
	of these variances as a function of the system size $n$, indicating the emergence of barren plateaus: for even moderately system sizes random parameters will most likely correspond to a vanishing-gradient region of the optimization landscape.
	Note that reducing the depth of the circuits and choosing a different cost can in principle alleviate barren plateaus for this specific example \cite{cerezo2020cost}. 
	Still, as we aim at testing \zinit\ in challenging situations, we did not resort to such alternatives.

\subsection{Individual optimizations}
\label{sec:appendix:sgd}

\begin{figure}
	\includegraphics[width=0.49\textwidth]{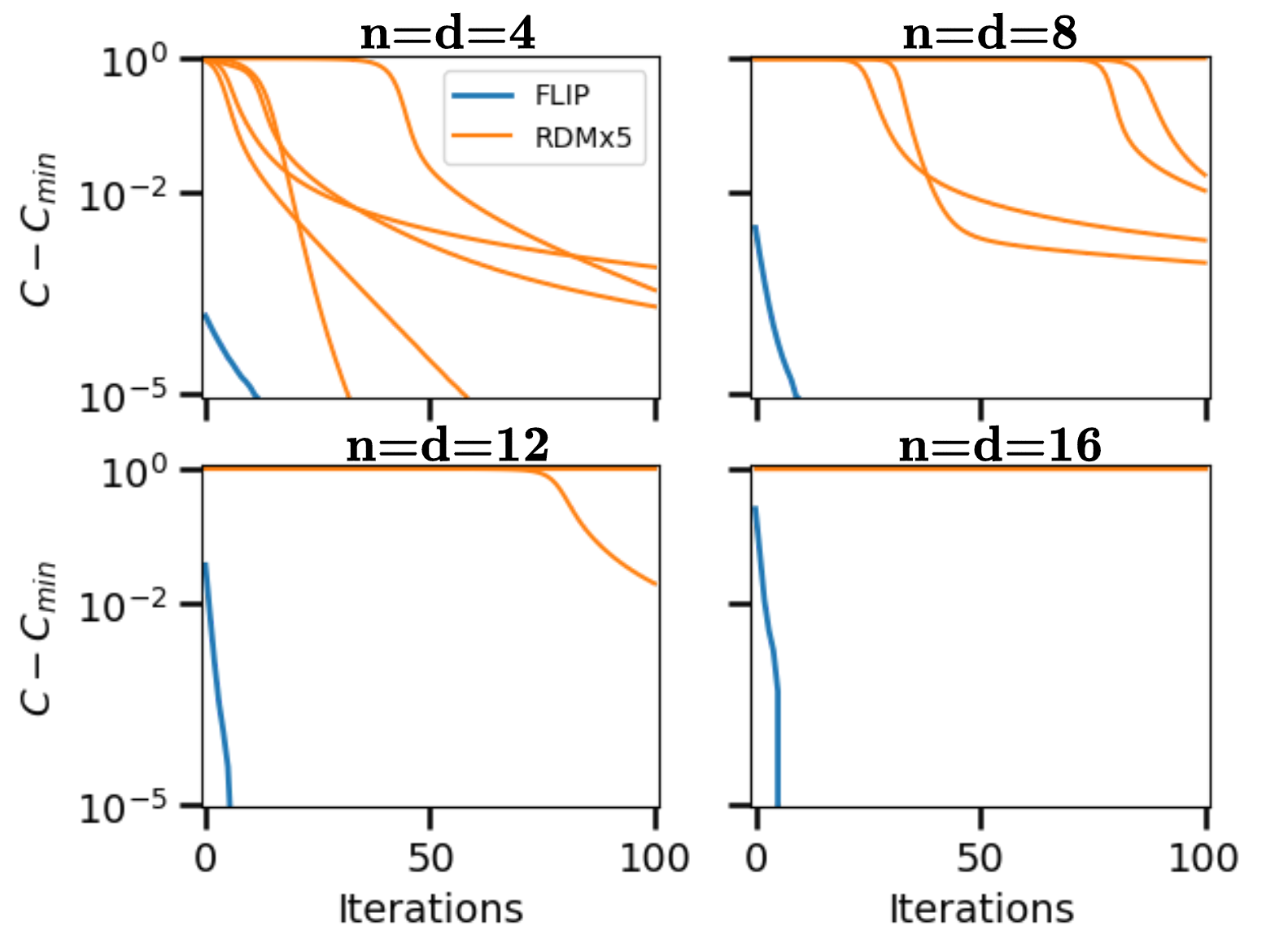}
	\caption{
		Individual optimization results for the state preparation problems with circuits of different sizes from $n=d=4$ to $n=d=16$. Optimizations starting with \zinit (blue curves) are compared to random initialization (orange curves). For each problem random initialization is repeated $5$ times.
	}
	\label{fig:barren_sgd}
\end{figure}

A selected subset of optimizations is plotted in Fig.~\ref{fig:barren_sgd}, spanning problems of different sizes from $n=4$ to $16$ qubits, with a circuit depth equal to the number of qubits, i.e., $n=d$. 
For each of these $4$ individual problems, the optimizations starting from random parameters are repeated $5$ times.

Looking at optimization traces for randomly initialized circuits, one can see that convergence quickly deteriorates when the size of the circuits involved is increased.
Already for system size $n=8$ qubits, only two out of the five runs show progress before c.$50$ iterations.
For the case $n=12$ qubits only one out of the five optimizations converged reasonably close to the minimum, while for $n=16$ qubits none of the runs manage to even slightly improve the objective value.  

In contrast, for all the problem instances, circuits initialized with \zinit\ converged quickly  with an advantage over random initialization which is most appreciable for the largest circuits considered. 
The case $n=d=16$ qubits corresponds to problems twice as deep and as wide (with four times more parameters) than the largest circuit seen during training.
This probes the ability of \zinit\ to correctly identify successful parameters patterns based on problems of moderate sizes which can be extrapolated to larger ones.

\subsection{Optimization with Adam}
\label{sec:appendix:adam}
\begin{figure}
	\includegraphics[width=0.48\textwidth]{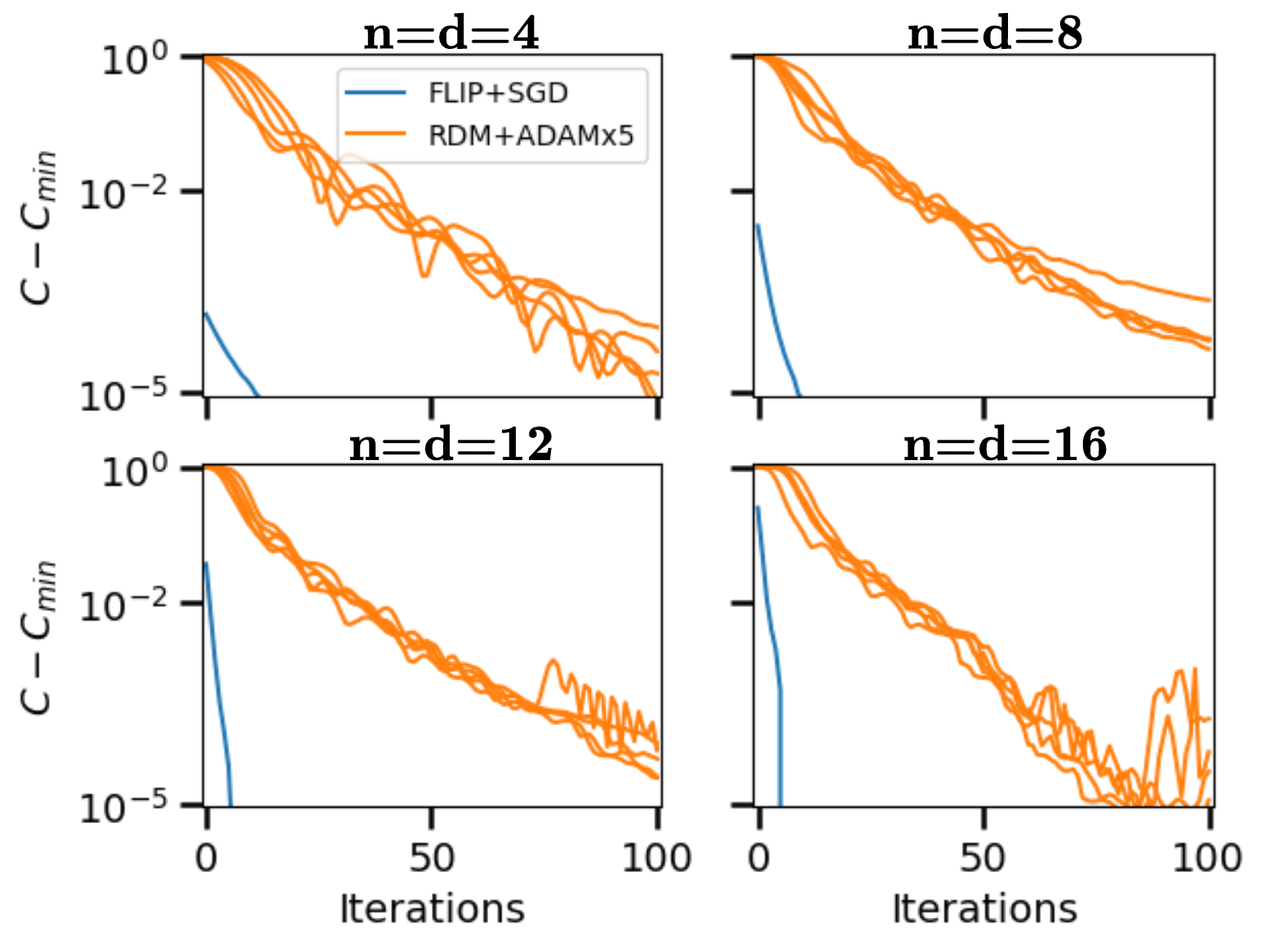}
	\caption{
    	Optimization with Adam.
    	Results similar to Fig.\ref{fig:barren_sgd}, but for randomly initialized circuits the gradient-descent method Adam is used.
	}
	\label{fig:barren_adam}
\end{figure}

Results presented in Fig.~\ref{fig:barren_sgd} were obtained with standard gradient descent, where the step in parameter space taken at each iteration is obtained by multiplying the gradient vector by a fixed scalar value (the learning rate). 
Many variations of this simple routine have been explored, especially in the context of machine learning.
In particular, Adam \cite{kingma2014adam} has become a popular optimizer used when training neural networks.
In Fig.~\ref{fig:barren_adam} we reproduce the results displayed in Fig.~\ref{fig:barren_sgd}, with the exception that optimizations starting from random parameters (green curves) are now performed using Adam.
In contrast to Fig.~\ref{fig:barren_sgd}, despite the presence of barren plateaus, optimizations with Adam are able to progress for any of the sizes studied.
Note that even in this case optimizations with circuits initialized with \zinit, followed by standard gradient descent, converge significantly faster.

Adam relies on adaptive learning rates, which can effectively be quite large especially in the first steps of optimizations~\cite{DBLP:conf/iclr/LiuJHCLG020}.
That is, in principle vanishing gradients could be magnified by re-scaling them by an arbitrarily large scalar value or tensor. 
However we expect that when noise is included and dominate the magnitude of these gradients (or of the higher order derivatives \cite{PhysRevA.103.012405}), 
such approach should fail as it will consist of taking large steps in almost random directions. 
In the next section, we verify that it is indeed the case for Adam.

\subsection{Optimization with noise}
\label{sec:appendix:noisy}
\begin{figure}
	\includegraphics[width=0.49\textwidth]{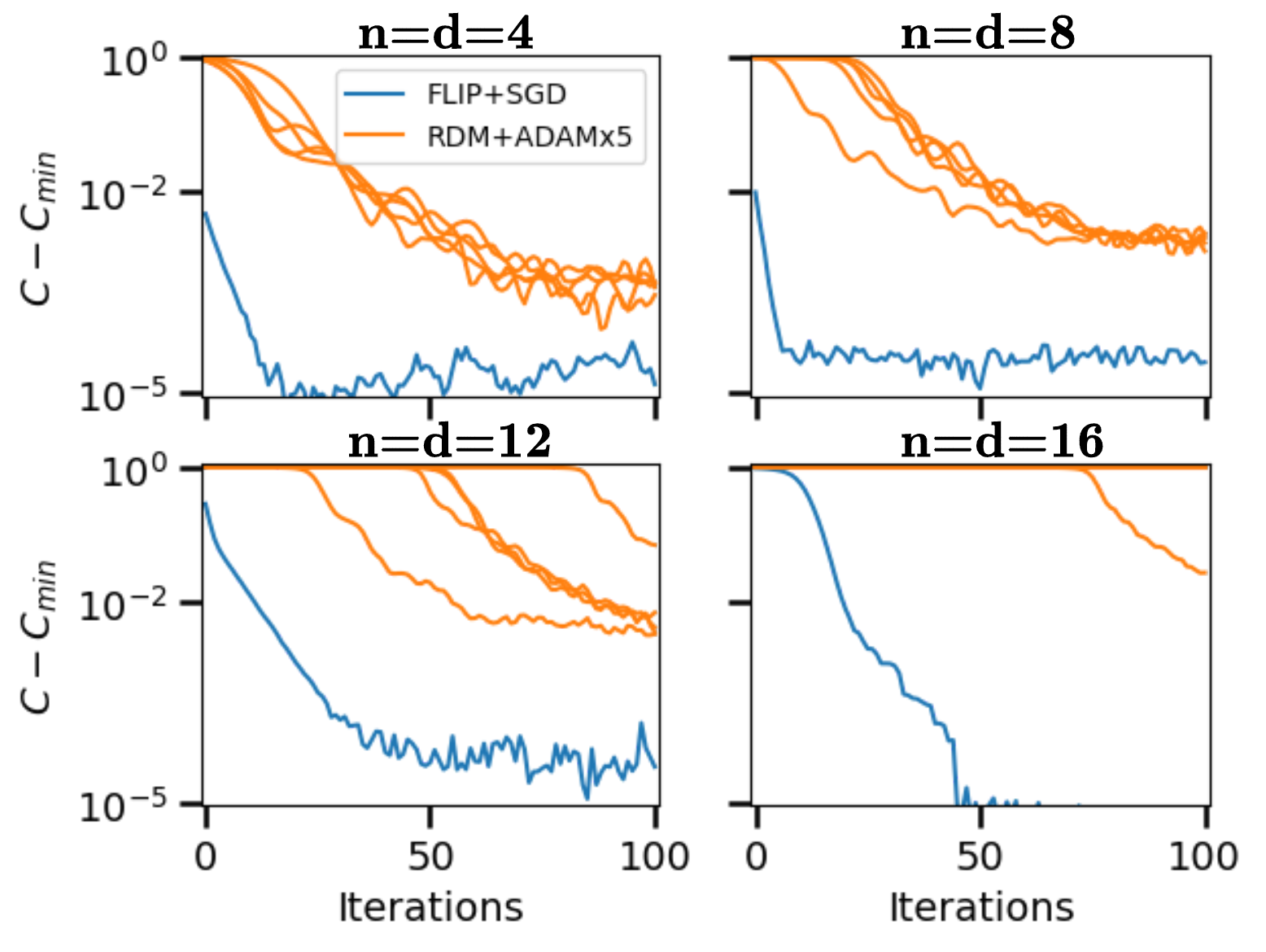}
	\caption{
    	Optimization with noisy gradients. 
    	Exact same optimizations as shown in Fig.\ref{fig:barren_adam}, except for the addition of noise in the gradients.
    	This noise is taken to be i.i.d Gaussian distributed with zero mean and a standard deviation of $\sigma_n=0.01$.
	This noise is included in the gradients during optimization and also when training \zinit.
	}
	\label{fig:barren_noise}
\end{figure}

So far, we have assumed noiseless gradients. 
However in practice these gradients are estimated through a finite number of measurements and statistical noise needs to be accounted for.
For that purpose we now include additive i.i.d Gaussian noise with standard deviation $\sigma_n=0.01$ and report results of optimization performed in this more realistic scenario in Fig.~\ref{fig:barren_noise}.  

For random initialization (green curves) we resort to Adam as it was found beneficial in the noiseless scenario.
Noticeably the ability to overcome barren plateaus seen in the noiseless case is now suppressed, as displayed in Fig.~\ref{fig:barren_noise} for the cases $n=d\geq 12$ (bottom row). 
This confirms our intuition that strategies relying on re-scaling gradients by large values are probably not viable in realistic conditions.

In the case of \zinit (blue curves), both its training and testing are subjected to noisy gradients. 
When compared to Fig.~\ref{fig:barren_adam}, one can notice a slight decrease in the overall rate of convergence. 
Still the main conclusions drawn for the noiseless case remain unchanged: (i) for circuit sizes pertaining to the training distributions circuits are initialized already close to the optimal parameters (ii) for larger circuits, despite starting further away, the parameters initialized with \zinit\ remains in a region where the gradient signal is strong enough such that the initial parameters can be quickly refined by gradient-descent.
As such, these results confirm the ability of \zinit\ to mitigate the barren plateau phenomenon as long as underlying patterns in the optimal parameters over varied circuit sizes and objectives can be learnt.

\section{Max-cut with QAOA}
\label{sec:supp:qaoa}
\begin{figure}
	\includegraphics[width=0.48\textwidth]{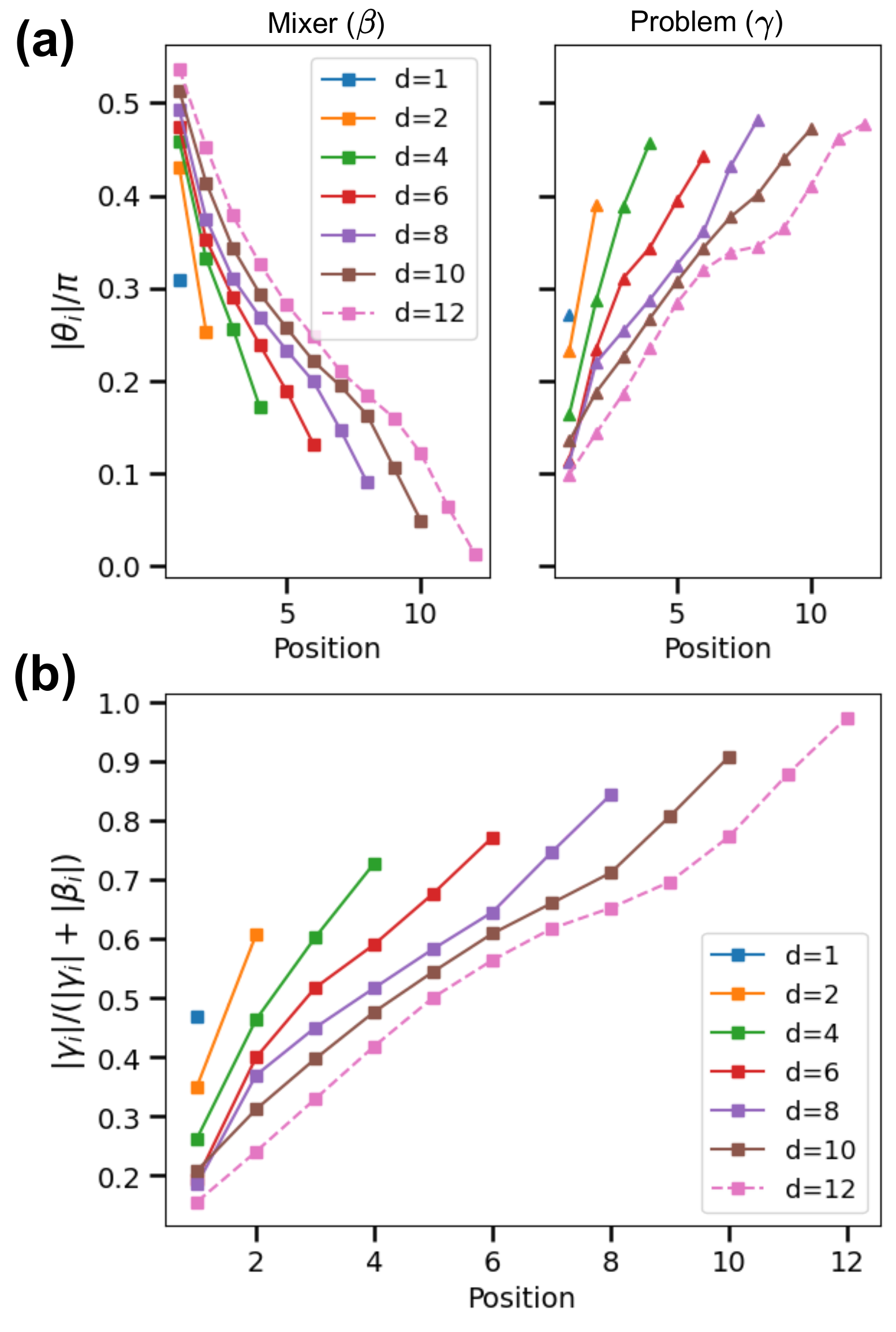}
	\caption{Patterns in the initial parameters learnt by \zinit\ for the max-cut problems with QAOA  circuits. 
	\textbf{(a)} Absolute values of the parameters $\theta_l$ for varied circuit sizes within $1 \leq d \leq 12$ layers (colours in legend) and varied layer position $l$ (x-axis). These are grouped by parameters pertaining to the mixer ($\beta_l$) and  problem ($\gamma_l$) unitaries, in the left and right panels respectively. The values (y-axis) are reported as a fraction of $\pi$. \textbf{(b)} Ratio $|\gamma_l|/(|\gamma_l| + |\beta_l|)$.
	}
	\label{fig:qaoa_patterns}
\end{figure}

Here we present the patterns in the initial parameters learnt by \zinit\ over the distribution of max-cut problems with the QAOA circuits discussed in Sec.~\ref{s:qaoa}.
We plot in Fig.~\ref{fig:qaoa_patterns} the initial parameter values, produced by \zinit, for circuits of dimension $d_{\taskidx}\in [1,12]$ layers (colours). 
The absolute values (rescaled by $\pi$) of the parameters $\beta_l$ and $\gamma_l$ are reported in Fig.~\ref{fig:qaoa_patterns}(a), left and right panels respectively, as a function of the layer position $l$. In Fig.~\ref{fig:qaoa_patterns}(b) we plot the ratio $|\gamma_l|/(|\gamma_l|+|\beta_l|$).

Clear patterns emerge: at fixed circuit depth the absolute values of $\beta_l$ ($\gamma_l$) decrease (increase) for increasing layer position $l$.
In addition, one can also inspect changes in the initial values of a fixed parameter but for circuits of varied sizes. In this case $\beta_l$ ($\gamma_l$) are also found to increase (decrease) for increasing circuit depths.
These patterns are qualitatively similar to the patterns discovered and exploited in the context of max-cut QAOA problems \cite{PhysRevX.10.021067} and are reminiscent of smooth schedules used in adiabatic quantum computing \cite{RevModPhys.90.015002}.

In \cite{PhysRevX.10.021067} the authors conducted a thorough study of optimal parameters found for different QAOA circuit sizes over many repeated optimizations.
Insights gained on the structure of these optimal parameters for small problems were subsequently exploited to engineer a sequential optimization strategy. 
A circuit is grown one layer at a time and the initial parameters of new circuits are obtained by extrapolating the optimized parameters of the previous circuit. Directly learning these patterns over many circuit sizes, as made possible using \zinit, avoids such sequential procedure and handcrafted extrapolation rules.

\section{Fermi-Hubbard model}
\label{sec:extra:fhm}
\begin{figure}
	\includegraphics[width=0.48\textwidth]{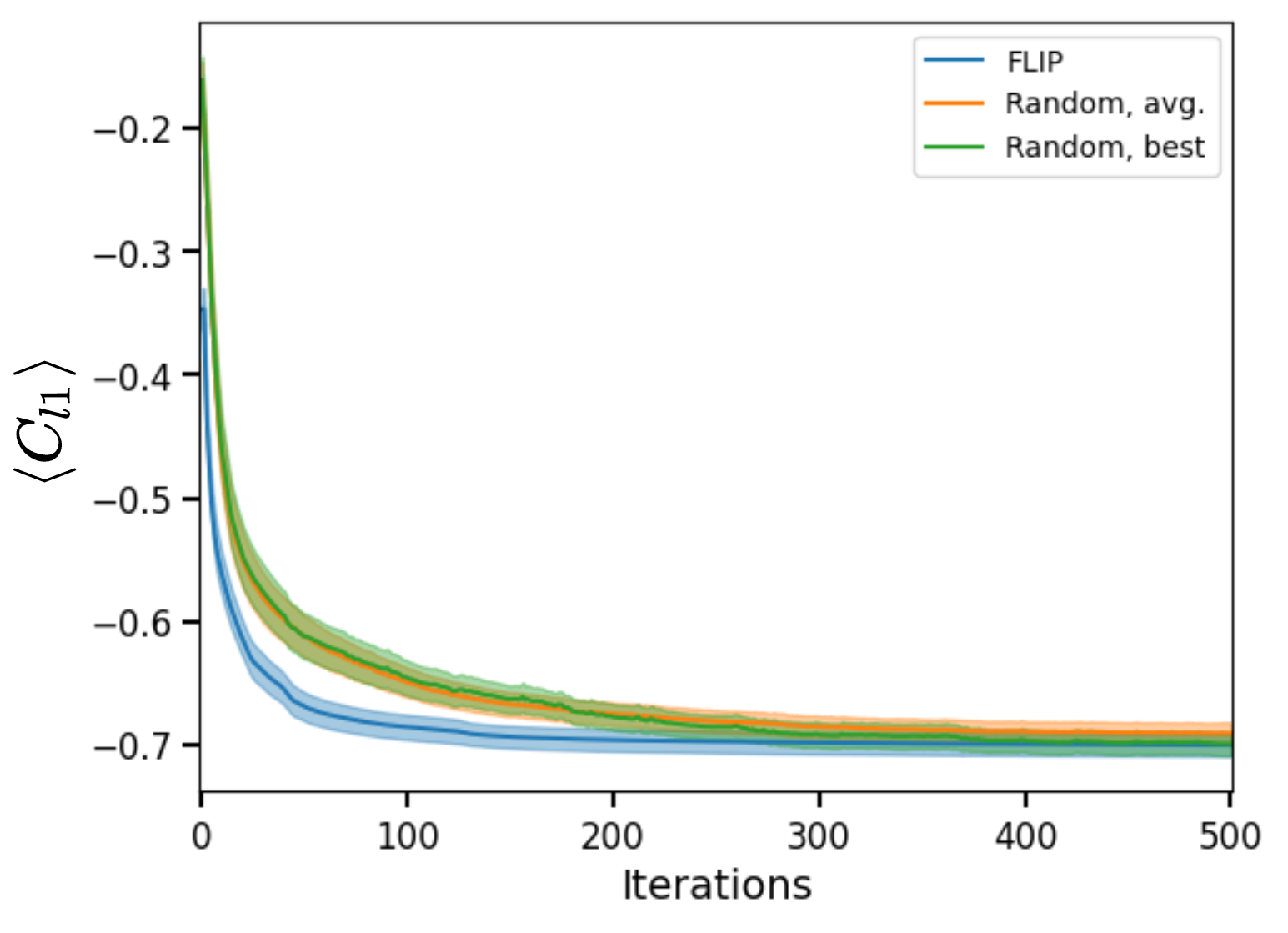}
	\caption{
	Fermi-Hubbard model with $500$ optimization steps. 
	Similar to Fig.~\ref{fig:fhm}, the average of the normalized costs for circuits initialized with \zinit\ and randomly are compared, but with a larger number of iterations ($500$ steps instead of $100$).
	}
	\label{fig:fhm500}
\end{figure}

\begin{figure}
	\includegraphics[width=0.48\textwidth]{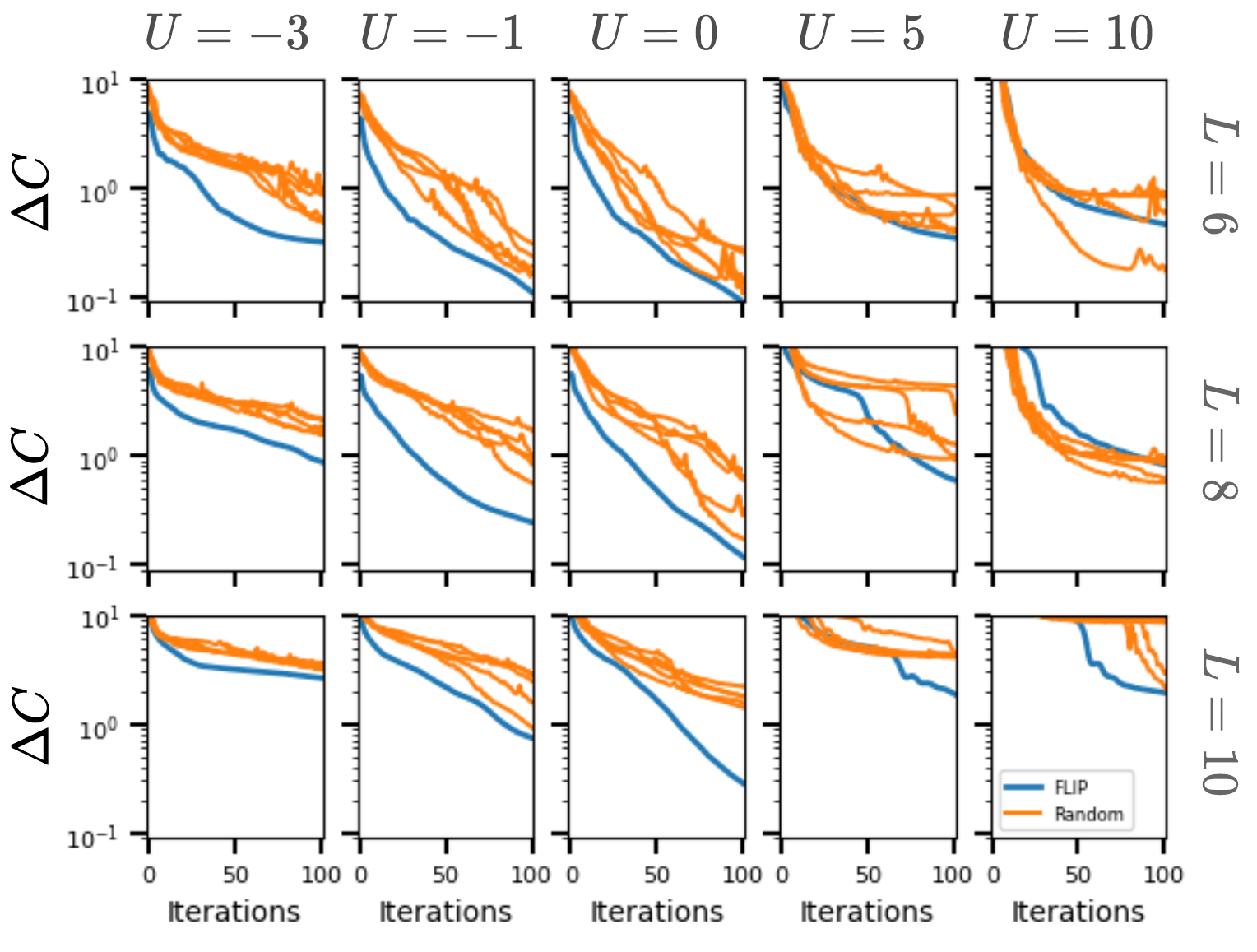}
	\caption{Fermi-Hubbard model with interaction strengths $U\in[-3,10]$. 
	Results of individual optimizations with randomly and \zinit-initialized circuits. 
	The training of \zinit\ is also performed based on this extended range of interaction values.
	}
	\label{fig:fhmNegU}
\end{figure}

In this appendix we provide additional results obtained for the Fermi-Hubbard model described in Sec.~\ref{s:fhm}.

In Fig. \ref{fig:fhm500} we report results of optimizations of circuits either initialized with \zinit\ or randomly with an extended number of optimization steps ($500$ instead of the $100$ iterations showcased in the main text). 
Similarly to Fig.~\ref{fig:fhm}(a) the values of the normalized costs are averaged over all the testing problems and displayed as a function of the number of iterations performed.
One can see that the gap between the different initializers shrinks only after a large number of steps, $c. 300$. 
Circuits initialized with \zinit\ converge faster and show (slightly) better final convergence in average, even when compared to the best out of $5$ random initializations per problem.

At last, we also consider the case where interaction strengths $U$ of the FHM can adopt both positive and negative values.
Results for the case $U \in [-3,10]$ are reported in Fig.~\ref{fig:fhmNegU} for individual testing problems spanning both positive and negative interaction cases.
Details of the training and testing are the same as in Sec.~\ref{s:fhm}, except for the interaction strength which is now drawn uniformly within its new range.
These results are in line with the ones reported in the main text: in almost all cases circuits initialized with \zinit\ converge to lower values than the best out of $5$ random initializations.  

However when considering the larger interval of interactions $U \in [-10,10]$, we found it harder to train \zinit\ and in some cases our strategy under-performs random initialization.
Inspecting the absolute values of the normalized costs (used for training) as a function of the interaction strengths, we notice that they were significantly smaller for the range $U=[-10,-3]$ than for the range of interaction strengths $U=[-3,10]$.
We believe that it explains the degradation in performance of \zinit\ and expect that a better choice of normalization could resolve these difficulties.
Additionally, it would be interesting to extend \zinit\ to also learn optimal learning rates \cite{li2017meta} to be used after initialization.
Similarly to the initial parameters, these could be made dependent on the underlying details of the problems, and in this case compensate for the changes in magnitudes of the cost for varied values of interactions.


\begin{thebibliography}{51}%
\makeatletter
\providecommand \@ifxundefined [1]{%
 \@ifx{#1\undefined}
}%
\providecommand \@ifnum [1]{%
 \ifnum #1\expandafter \@firstoftwo
 \else \expandafter \@secondoftwo
 \fi
}%
\providecommand \@ifx [1]{%
 \ifx #1\expandafter \@firstoftwo
 \else \expandafter \@secondoftwo
 \fi
}%
\providecommand \natexlab [1]{#1}%
\providecommand \enquote  [1]{``#1''}%
\providecommand \bibnamefont  [1]{#1}%
\providecommand \bibfnamefont [1]{#1}%
\providecommand \citenamefont [1]{#1}%
\providecommand \href@noop [0]{\@secondoftwo}%
\providecommand \href [0]{\begingroup \@sanitize@url \@href}%
\providecommand \@href[1]{\@@startlink{#1}\@@href}%
\providecommand \@@href[1]{\endgroup#1\@@endlink}%
\providecommand \@sanitize@url [0]{\catcode `\\12\catcode `\$12\catcode
  `\&12\catcode `\#12\catcode `\^12\catcode `\_12\catcode `\%12\relax}%
\providecommand \@@startlink[1]{}%
\providecommand \@@endlink[0]{}%
\providecommand \url  [0]{\begingroup\@sanitize@url \@url }%
\providecommand \@url [1]{\endgroup\@href {#1}{\urlprefix }}%
\providecommand \urlprefix  [0]{URL }%
\providecommand \Eprint [0]{\href }%
\providecommand \doibase [0]{https://doi.org/}%
\providecommand \selectlanguage [0]{\@gobble}%
\providecommand \bibinfo  [0]{\@secondoftwo}%
\providecommand \bibfield  [0]{\@secondoftwo}%
\providecommand \translation [1]{[#1]}%
\providecommand \BibitemOpen [0]{}%
\providecommand \bibitemStop [0]{}%
\providecommand \bibitemNoStop [0]{.\EOS\space}%
\providecommand \EOS [0]{\spacefactor3000\relax}%
\providecommand \BibitemShut  [1]{\csname bibitem#1\endcsname}%
\let\auto@bib@innerbib\@empty
%</preamble>
\bibitem [{\citenamefont {{Cerezo}}\ \emph {et~al.}(2020)\citenamefont
  {{Cerezo}}, \citenamefont {{Arrasmith}}, \citenamefont {{Babbush}},
  \citenamefont {{Benjamin}}, \citenamefont {{Endo}}, \citenamefont {{Fujii}},
  \citenamefont {{McClean}}, \citenamefont {{Mitarai}}, \citenamefont {{Yuan}},
  \citenamefont {{Cincio}},\ and\ \citenamefont
  {{Coles}}}]{cerezo2020variational}%
  \BibitemOpen
  \bibfield  {author} {\bibinfo {author} {\bibfnamefont {M.}~\bibnamefont
  {{Cerezo}}}, \bibinfo {author} {\bibfnamefont {A.}~\bibnamefont
  {{Arrasmith}}}, \bibinfo {author} {\bibfnamefont {R.}~\bibnamefont
  {{Babbush}}}, \bibinfo {author} {\bibfnamefont {S.~C.}\ \bibnamefont
  {{Benjamin}}}, \bibinfo {author} {\bibfnamefont {S.}~\bibnamefont {{Endo}}},
  \bibinfo {author} {\bibfnamefont {K.}~\bibnamefont {{Fujii}}}, \bibinfo
  {author} {\bibfnamefont {J.~R.}\ \bibnamefont {{McClean}}}, \bibinfo {author}
  {\bibfnamefont {K.}~\bibnamefont {{Mitarai}}}, \bibinfo {author}
  {\bibfnamefont {X.}~\bibnamefont {{Yuan}}}, \bibinfo {author} {\bibfnamefont
  {L.}~\bibnamefont {{Cincio}}},\ and\ \bibinfo {author} {\bibfnamefont
  {P.~J.}\ \bibnamefont {{Coles}}},\ }\bibfield  {title} {\bibinfo {title}
  {{Variational Quantum Algorithms}},\ }\href {http://arxiv.org/abs/2012.09265}
  {\bibfield  {journal} {\bibinfo  {journal} {arXiv:2012.09265}\ } (\bibinfo
  {year} {2020})}\BibitemShut {NoStop}%
\bibitem [{\citenamefont {Bharti}\ \emph {et~al.}(2021)\citenamefont {Bharti},
  \citenamefont {Cervera-Lierta}, \citenamefont {Kyaw}, \citenamefont {Haug},
  \citenamefont {Alperin-Lea}, \citenamefont {Anand}, \citenamefont {Degroote},
  \citenamefont {Heimonen}, \citenamefont {Kottmann}, \citenamefont {Menke}
  \emph {et~al.}}]{bharti2021noisy}%
  \BibitemOpen
  \bibfield  {author} {\bibinfo {author} {\bibfnamefont {K.}~\bibnamefont
  {Bharti}}, \bibinfo {author} {\bibfnamefont {A.}~\bibnamefont
  {Cervera-Lierta}}, \bibinfo {author} {\bibfnamefont {T.~H.}\ \bibnamefont
  {Kyaw}}, \bibinfo {author} {\bibfnamefont {T.}~\bibnamefont {Haug}}, \bibinfo
  {author} {\bibfnamefont {S.}~\bibnamefont {Alperin-Lea}}, \bibinfo {author}
  {\bibfnamefont {A.}~\bibnamefont {Anand}}, \bibinfo {author} {\bibfnamefont
  {M.}~\bibnamefont {Degroote}}, \bibinfo {author} {\bibfnamefont
  {H.}~\bibnamefont {Heimonen}}, \bibinfo {author} {\bibfnamefont {J.~S.}\
  \bibnamefont {Kottmann}}, \bibinfo {author} {\bibfnamefont {T.}~\bibnamefont
  {Menke}}, \emph {et~al.},\ }\bibfield  {title} {\bibinfo {title} {Noisy
  intermediate-scale quantum (nisq) algorithms},\ }\href
  {http://arxiv.org/abs/2101.08448} {\bibfield  {journal} {\bibinfo  {journal}
  {arXiv:2101.08448}\ } (\bibinfo {year} {2021})}\BibitemShut {NoStop}%
\bibitem [{\citenamefont {Benedetti}\ \emph
  {et~al.}(2019{\natexlab{a}})\citenamefont {Benedetti}, \citenamefont {Lloyd},
  \citenamefont {Sack},\ and\ \citenamefont
  {Fiorentini}}]{benedetti2019parameterized}%
  \BibitemOpen
  \bibfield  {author} {\bibinfo {author} {\bibfnamefont {M.}~\bibnamefont
  {Benedetti}}, \bibinfo {author} {\bibfnamefont {E.}~\bibnamefont {Lloyd}},
  \bibinfo {author} {\bibfnamefont {S.}~\bibnamefont {Sack}},\ and\ \bibinfo
  {author} {\bibfnamefont {M.}~\bibnamefont {Fiorentini}},\ }\bibfield  {title}
  {\bibinfo {title} {Parameterized quantum circuits as machine learning
  models},\ }\href {https://doi.org/10.1088/2058-9565/ab4eb5} {\bibfield
  {journal} {\bibinfo  {journal} {Quantum Sci. Technol.}\ }\textbf {\bibinfo
  {volume} {4}},\ \bibinfo {pages} {043001} (\bibinfo {year}
  {2019}{\natexlab{a}})}\BibitemShut {NoStop}%
\bibitem [{\citenamefont {McClean}\ \emph {et~al.}(2018)\citenamefont
  {McClean}, \citenamefont {Boixo}, \citenamefont {Smelyanskiy}, \citenamefont
  {Babbush},\ and\ \citenamefont {Neven}}]{McClean2018}%
  \BibitemOpen
  \bibfield  {author} {\bibinfo {author} {\bibfnamefont {J.~R.}\ \bibnamefont
  {McClean}}, \bibinfo {author} {\bibfnamefont {S.}~\bibnamefont {Boixo}},
  \bibinfo {author} {\bibfnamefont {V.~N.}\ \bibnamefont {Smelyanskiy}},
  \bibinfo {author} {\bibfnamefont {R.}~\bibnamefont {Babbush}},\ and\ \bibinfo
  {author} {\bibfnamefont {H.}~\bibnamefont {Neven}},\ }\bibfield  {title}
  {\bibinfo {title} {Barren plateaus in quantum neural network training
  landscapes},\ }\href {https://doi.org/10.1038/s41467-018-07090-4} {\bibfield
  {journal} {\bibinfo  {journal} {Nat. Commun.}\ }\textbf {\bibinfo {volume}
  {9}},\ \bibinfo {pages} {4812} (\bibinfo {year} {2018})}\BibitemShut
  {NoStop}%
\bibitem [{\citenamefont {Cerezo}\ \emph {et~al.}(2021)\citenamefont {Cerezo},
  \citenamefont {Sone}, \citenamefont {Volkoff}, \citenamefont {Cincio},\ and\
  \citenamefont {Coles}}]{cerezo2020cost}%
  \BibitemOpen
  \bibfield  {author} {\bibinfo {author} {\bibfnamefont {M.}~\bibnamefont
  {Cerezo}}, \bibinfo {author} {\bibfnamefont {A.}~\bibnamefont {Sone}},
  \bibinfo {author} {\bibfnamefont {T.}~\bibnamefont {Volkoff}}, \bibinfo
  {author} {\bibfnamefont {L.}~\bibnamefont {Cincio}},\ and\ \bibinfo {author}
  {\bibfnamefont {P.~J.}\ \bibnamefont {Coles}},\ }\bibfield  {title} {\bibinfo
  {title} {Cost function dependent barren plateaus in shallow parametrized
  quantum circuits},\ }\href {https://doi.org/10.1038/s41467-021-21728-w}
  {\bibfield  {journal} {\bibinfo  {journal} {Nat. Commun.}\ }\textbf {\bibinfo
  {volume} {12}},\ \bibinfo {pages} {1791} (\bibinfo {year}
  {2021})}\BibitemShut {NoStop}%
\bibitem [{\citenamefont {Zhou}\ \emph {et~al.}(2020)\citenamefont {Zhou},
  \citenamefont {Wang}, \citenamefont {Choi}, \citenamefont {Pichler},\ and\
  \citenamefont {Lukin}}]{PhysRevX.10.021067}%
  \BibitemOpen
  \bibfield  {author} {\bibinfo {author} {\bibfnamefont {L.}~\bibnamefont
  {Zhou}}, \bibinfo {author} {\bibfnamefont {S.-T.}\ \bibnamefont {Wang}},
  \bibinfo {author} {\bibfnamefont {S.}~\bibnamefont {Choi}}, \bibinfo {author}
  {\bibfnamefont {H.}~\bibnamefont {Pichler}},\ and\ \bibinfo {author}
  {\bibfnamefont {M.~D.}\ \bibnamefont {Lukin}},\ }\bibfield  {title} {\bibinfo
  {title} {Quantum approximate optimization algorithm: Performance, mechanism,
  and implementation on near-term devices},\ }\href
  {https://doi.org/10.1103/PhysRevX.10.021067} {\bibfield  {journal} {\bibinfo
  {journal} {Phys. Rev. X}\ }\textbf {\bibinfo {volume} {10}},\ \bibinfo
  {pages} {021067} (\bibinfo {year} {2020})}\BibitemShut {NoStop}%
\bibitem [{\citenamefont {Schuld}\ \emph {et~al.}(2019)\citenamefont {Schuld},
  \citenamefont {Bergholm}, \citenamefont {Gogolin}, \citenamefont {Izaac},\
  and\ \citenamefont {Killoran}}]{PhysRevA.99.032331}%
  \BibitemOpen
  \bibfield  {author} {\bibinfo {author} {\bibfnamefont {M.}~\bibnamefont
  {Schuld}}, \bibinfo {author} {\bibfnamefont {V.}~\bibnamefont {Bergholm}},
  \bibinfo {author} {\bibfnamefont {C.}~\bibnamefont {Gogolin}}, \bibinfo
  {author} {\bibfnamefont {J.}~\bibnamefont {Izaac}},\ and\ \bibinfo {author}
  {\bibfnamefont {N.}~\bibnamefont {Killoran}},\ }\bibfield  {title} {\bibinfo
  {title} {Evaluating analytic gradients on quantum hardware},\ }\href
  {https://doi.org/10.1103/PhysRevA.99.032331} {\bibfield  {journal} {\bibinfo
  {journal} {Phys. Rev. A}\ }\textbf {\bibinfo {volume} {99}},\ \bibinfo
  {pages} {032331} (\bibinfo {year} {2019})}\BibitemShut {NoStop}%
\bibitem [{\citenamefont {Grant}\ \emph {et~al.}(2019)\citenamefont {Grant},
  \citenamefont {Wossnig}, \citenamefont {Ostaszewski},\ and\ \citenamefont
  {Benedetti}}]{Grant2019initialization}%
  \BibitemOpen
  \bibfield  {author} {\bibinfo {author} {\bibfnamefont {E.}~\bibnamefont
  {Grant}}, \bibinfo {author} {\bibfnamefont {L.}~\bibnamefont {Wossnig}},
  \bibinfo {author} {\bibfnamefont {M.}~\bibnamefont {Ostaszewski}},\ and\
  \bibinfo {author} {\bibfnamefont {M.}~\bibnamefont {Benedetti}},\ }\bibfield
  {title} {\bibinfo {title} {An initialization strategy for addressing barren
  plateaus in parametrized quantum circuits},\ }\href
  {https://doi.org/10.22331/q-2019-12-09-214} {\bibfield  {journal} {\bibinfo
  {journal} {{Quantum}}\ }\textbf {\bibinfo {volume} {3}},\ \bibinfo {pages}
  {214} (\bibinfo {year} {2019})}\BibitemShut {NoStop}%
\bibitem [{\citenamefont {Grimsley}\ \emph {et~al.}(2019)\citenamefont
  {Grimsley}, \citenamefont {Economou}, \citenamefont {Barnes},\ and\
  \citenamefont {Mayhall}}]{Grimsley2019}%
  \BibitemOpen
  \bibfield  {author} {\bibinfo {author} {\bibfnamefont {H.~R.}\ \bibnamefont
  {Grimsley}}, \bibinfo {author} {\bibfnamefont {S.~E.}\ \bibnamefont
  {Economou}}, \bibinfo {author} {\bibfnamefont {E.}~\bibnamefont {Barnes}},\
  and\ \bibinfo {author} {\bibfnamefont {N.~J.}\ \bibnamefont {Mayhall}},\
  }\bibfield  {title} {\bibinfo {title} {An adaptive variational algorithm for
  exact molecular simulations on a quantum computer},\ }\href
  {https://doi.org/10.1038/s41467-019-10988-2} {\bibfield  {journal} {\bibinfo
  {journal} {Nat. Commun.}\ }\textbf {\bibinfo {volume} {10}},\ \bibinfo
  {pages} {3007} (\bibinfo {year} {2019})}\BibitemShut {NoStop}%
\bibitem [{\citenamefont {Verdon}\ \emph {et~al.}(2019)\citenamefont {Verdon},
  \citenamefont {Broughton}, \citenamefont {McClean}, \citenamefont {Sung},
  \citenamefont {Babbush}, \citenamefont {Jiang}, \citenamefont {Neven},\ and\
  \citenamefont {Mohseni}}]{verdon2019learning}%
  \BibitemOpen
  \bibfield  {author} {\bibinfo {author} {\bibfnamefont {G.}~\bibnamefont
  {Verdon}}, \bibinfo {author} {\bibfnamefont {M.}~\bibnamefont {Broughton}},
  \bibinfo {author} {\bibfnamefont {J.~R.}\ \bibnamefont {McClean}}, \bibinfo
  {author} {\bibfnamefont {K.~J.}\ \bibnamefont {Sung}}, \bibinfo {author}
  {\bibfnamefont {R.}~\bibnamefont {Babbush}}, \bibinfo {author} {\bibfnamefont
  {Z.}~\bibnamefont {Jiang}}, \bibinfo {author} {\bibfnamefont
  {H.}~\bibnamefont {Neven}},\ and\ \bibinfo {author} {\bibfnamefont
  {M.}~\bibnamefont {Mohseni}},\ }\bibfield  {title} {\bibinfo {title}
  {Learning to learn with quantum neural networks via classical neural
  networks},\ }\href {http://arxiv.org/abs/1907.05415} {\bibfield  {journal}
  {\bibinfo  {journal} {arXiv:1907.05415}\ } (\bibinfo {year}
  {2019})}\BibitemShut {NoStop}%
\bibitem [{\citenamefont {Wilson}\ \emph {et~al.}(2019)\citenamefont {Wilson},
  \citenamefont {Stromswold}, \citenamefont {Wudarski}, \citenamefont
  {Hadfield}, \citenamefont {Tubman},\ and\ \citenamefont
  {Rieffel}}]{wilson2019optimizing}%
  \BibitemOpen
  \bibfield  {author} {\bibinfo {author} {\bibfnamefont {M.}~\bibnamefont
  {Wilson}}, \bibinfo {author} {\bibfnamefont {S.}~\bibnamefont {Stromswold}},
  \bibinfo {author} {\bibfnamefont {F.}~\bibnamefont {Wudarski}}, \bibinfo
  {author} {\bibfnamefont {S.}~\bibnamefont {Hadfield}}, \bibinfo {author}
  {\bibfnamefont {N.~M.}\ \bibnamefont {Tubman}},\ and\ \bibinfo {author}
  {\bibfnamefont {E.}~\bibnamefont {Rieffel}},\ }\bibfield  {title} {\bibinfo
  {title} {Optimizing quantum heuristics with meta-learning},\ }\href
  {http://arxiv.org/abs/1908.03185} {\bibfield  {journal} {\bibinfo  {journal}
  {arXiv:1908.03185}\ } (\bibinfo {year} {2019})}\BibitemShut {NoStop}%
\bibitem [{\citenamefont {Skolik}\ \emph {et~al.}(2021)\citenamefont {Skolik},
  \citenamefont {McClean}, \citenamefont {Mohseni}, \citenamefont {van~der
  Smagt},\ and\ \citenamefont {Leib}}]{skolik2020layerwise}%
  \BibitemOpen
  \bibfield  {author} {\bibinfo {author} {\bibfnamefont {A.}~\bibnamefont
  {Skolik}}, \bibinfo {author} {\bibfnamefont {J.~R.}\ \bibnamefont {McClean}},
  \bibinfo {author} {\bibfnamefont {M.}~\bibnamefont {Mohseni}}, \bibinfo
  {author} {\bibfnamefont {P.}~\bibnamefont {van~der Smagt}},\ and\ \bibinfo
  {author} {\bibfnamefont {M.}~\bibnamefont {Leib}},\ }\bibfield  {title}
  {\bibinfo {title} {Layerwise learning for quantum neural networks},\ }\href
  {https://doi.org/10.1007/s42484-020-00036-4} {\bibfield  {journal} {\bibinfo
  {journal} {Quantum Machine Intelligence}\ }\textbf {\bibinfo {volume} {3}},\
  \bibinfo {pages} {5} (\bibinfo {year} {2021})}\BibitemShut {NoStop}%
\bibitem [{\citenamefont {Finn}\ \emph {et~al.}(2017)\citenamefont {Finn},
  \citenamefont {Abbeel},\ and\ \citenamefont {Levine}}]{finn2017model}%
  \BibitemOpen
  \bibfield  {author} {\bibinfo {author} {\bibfnamefont {C.}~\bibnamefont
  {Finn}}, \bibinfo {author} {\bibfnamefont {P.}~\bibnamefont {Abbeel}},\ and\
  \bibinfo {author} {\bibfnamefont {S.}~\bibnamefont {Levine}},\ }\bibfield
  {title} {\bibinfo {title} {Model-agnostic meta-learning for fast adaptation
  of deep networks},\ }in\ \href
  {http://proceedings.mlr.press/v70/finn17a.html} {\emph {\bibinfo {booktitle}
  {Proceedings of the 34th International Conference on Machine Learning}}},\
  Vol.~\bibinfo {volume} {70}\ (\bibinfo  {publisher} {PMLR},\ \bibinfo {year}
  {2017})\BibitemShut {NoStop}%
\bibitem [{\citenamefont {Cervera-Lierta}\ \emph {et~al.}(2020)\citenamefont
  {Cervera-Lierta}, \citenamefont {Kottmann},\ and\ \citenamefont
  {Aspuru-Guzik}}]{cervera2020meta}%
  \BibitemOpen
  \bibfield  {author} {\bibinfo {author} {\bibfnamefont {A.}~\bibnamefont
  {Cervera-Lierta}}, \bibinfo {author} {\bibfnamefont {J.~S.}\ \bibnamefont
  {Kottmann}},\ and\ \bibinfo {author} {\bibfnamefont {A.}~\bibnamefont
  {Aspuru-Guzik}},\ }\bibfield  {title} {\bibinfo {title} {The meta-variational
  quantum eigensolver (meta-vqe): Learning energy profiles of parameterized
  hamiltonians for quantum simulation},\ }\href
  {http://arxiv.org/abs/2009.13545} {\bibfield  {journal} {\bibinfo  {journal}
  {arXiv:2009.13545}\ } (\bibinfo {year} {2020})}\BibitemShut {NoStop}%
\bibitem [{\citenamefont {Li}\ \emph {et~al.}(2017)\citenamefont {Li},
  \citenamefont {Zhou}, \citenamefont {Chen},\ and\ \citenamefont
  {Li}}]{li2017meta}%
  \BibitemOpen
  \bibfield  {author} {\bibinfo {author} {\bibfnamefont {Z.}~\bibnamefont
  {Li}}, \bibinfo {author} {\bibfnamefont {F.}~\bibnamefont {Zhou}}, \bibinfo
  {author} {\bibfnamefont {F.}~\bibnamefont {Chen}},\ and\ \bibinfo {author}
  {\bibfnamefont {H.}~\bibnamefont {Li}},\ }\bibfield  {title} {\bibinfo
  {title} {Meta-sgd: Learning to learn quickly for few-shot learning},\ }\href
  {http://arxiv.org/abs/1707.09835} {\bibfield  {journal} {\bibinfo  {journal}
  {arXiv:1707.09835}\ } (\bibinfo {year} {2017})}\BibitemShut {NoStop}%
\bibitem [{\citenamefont {Nichol}\ \emph {et~al.}(2018)\citenamefont {Nichol},
  \citenamefont {Achiam},\ and\ \citenamefont {Schulman}}]{nichol2018first}%
  \BibitemOpen
  \bibfield  {author} {\bibinfo {author} {\bibfnamefont {A.}~\bibnamefont
  {Nichol}}, \bibinfo {author} {\bibfnamefont {J.}~\bibnamefont {Achiam}},\
  and\ \bibinfo {author} {\bibfnamefont {J.}~\bibnamefont {Schulman}},\
  }\bibfield  {title} {\bibinfo {title} {On first-order meta-learning
  algorithms},\ }\href {http://arxiv.org/abs/1803.02999} {\bibfield  {journal}
  {\bibinfo  {journal} {arXiv:1803.02999}\ } (\bibinfo {year}
  {2018})}\BibitemShut {NoStop}%
\bibitem [{\citenamefont {Rusu}\ \emph {et~al.}(2018)\citenamefont {Rusu},
  \citenamefont {Rao}, \citenamefont {Sygnowski}, \citenamefont {Vinyals},
  \citenamefont {Pascanu}, \citenamefont {Osindero},\ and\ \citenamefont
  {Hadsell}}]{rusu2018meta}%
  \BibitemOpen
  \bibfield  {author} {\bibinfo {author} {\bibfnamefont {A.~A.}\ \bibnamefont
  {Rusu}}, \bibinfo {author} {\bibfnamefont {D.}~\bibnamefont {Rao}}, \bibinfo
  {author} {\bibfnamefont {J.}~\bibnamefont {Sygnowski}}, \bibinfo {author}
  {\bibfnamefont {O.}~\bibnamefont {Vinyals}}, \bibinfo {author} {\bibfnamefont
  {R.}~\bibnamefont {Pascanu}}, \bibinfo {author} {\bibfnamefont
  {S.}~\bibnamefont {Osindero}},\ and\ \bibinfo {author} {\bibfnamefont
  {R.}~\bibnamefont {Hadsell}},\ }\bibfield  {title} {\bibinfo {title}
  {Meta-learning with latent embedding optimization},\ }\href
  {http://arxiv.org/abs/1807.05960} {\bibfield  {journal} {\bibinfo  {journal}
  {arXiv:1807.05960}\ } (\bibinfo {year} {2018})}\BibitemShut {NoStop}%
\bibitem [{\citenamefont {Wecker}\ \emph {et~al.}(2015)\citenamefont {Wecker},
  \citenamefont {Hastings},\ and\ \citenamefont {Troyer}}]{Wecker_PRA2015}%
  \BibitemOpen
  \bibfield  {author} {\bibinfo {author} {\bibfnamefont {D.}~\bibnamefont
  {Wecker}}, \bibinfo {author} {\bibfnamefont {M.~B.}\ \bibnamefont
  {Hastings}},\ and\ \bibinfo {author} {\bibfnamefont {M.}~\bibnamefont
  {Troyer}},\ }\bibfield  {title} {\bibinfo {title} {Progress towards practical
  quantum variational algorithms},\ }\href
  {https://doi.org/10.1103/PhysRevA.92.042303} {\bibfield  {journal} {\bibinfo
  {journal} {Phys. Rev. A}\ }\textbf {\bibinfo {volume} {92}},\ \bibinfo
  {pages} {042303} (\bibinfo {year} {2015})}\BibitemShut {NoStop}%
\bibitem [{\citenamefont {McClean}\ \emph {et~al.}(2016)\citenamefont
  {McClean}, \citenamefont {Romero}, \citenamefont {Babbush},\ and\
  \citenamefont {Aspuru-Guzik}}]{McLean2016}%
  \BibitemOpen
  \bibfield  {author} {\bibinfo {author} {\bibfnamefont {J.~R.}\ \bibnamefont
  {McClean}}, \bibinfo {author} {\bibfnamefont {J.}~\bibnamefont {Romero}},
  \bibinfo {author} {\bibfnamefont {R.}~\bibnamefont {Babbush}},\ and\ \bibinfo
  {author} {\bibfnamefont {A.}~\bibnamefont {Aspuru-Guzik}},\ }\bibfield
  {title} {\bibinfo {title} {The theory of variational hybrid quantum-classical
  algorithms},\ }\href {http://stacks.iop.org/1367-2630/18/i=2/a=023023}
  {\bibfield  {journal} {\bibinfo  {journal} {New J. Phys.}\ }\textbf {\bibinfo
  {volume} {18}},\ \bibinfo {pages} {023023} (\bibinfo {year}
  {2016})}\BibitemShut {NoStop}%
\bibitem [{\citenamefont {Benedetti}\ \emph
  {et~al.}(2019{\natexlab{b}})\citenamefont {Benedetti}, \citenamefont
  {Garcia-Pintos}, \citenamefont {Perdomo}, \citenamefont {Leyton-Ortega},
  \citenamefont {Nam},\ and\ \citenamefont {Perdomo-Ortiz}}]{Benedetti2019}%
  \BibitemOpen
  \bibfield  {author} {\bibinfo {author} {\bibfnamefont {M.}~\bibnamefont
  {Benedetti}}, \bibinfo {author} {\bibfnamefont {D.}~\bibnamefont
  {Garcia-Pintos}}, \bibinfo {author} {\bibfnamefont {O.}~\bibnamefont
  {Perdomo}}, \bibinfo {author} {\bibfnamefont {V.}~\bibnamefont
  {Leyton-Ortega}}, \bibinfo {author} {\bibfnamefont {Y.}~\bibnamefont {Nam}},\
  and\ \bibinfo {author} {\bibfnamefont {A.}~\bibnamefont {Perdomo-Ortiz}},\
  }\bibfield  {title} {\bibinfo {title} {A generative modeling approach for
  benchmarking and training shallow quantum circuits},\ }\href
  {https://doi.org/10.1038/s41534-019-0157-8} {\bibfield  {journal} {\bibinfo
  {journal} {npj Quantum Inf.}\ }\textbf {\bibinfo {volume} {5}},\ \bibinfo
  {pages} {45} (\bibinfo {year} {2019}{\natexlab{b}})}\BibitemShut {NoStop}%
\bibitem [{\citenamefont {Liu}\ and\ \citenamefont
  {Wang}(2018)}]{liu2018differentiable}%
  \BibitemOpen
  \bibfield  {author} {\bibinfo {author} {\bibfnamefont {J.-G.}\ \bibnamefont
  {Liu}}\ and\ \bibinfo {author} {\bibfnamefont {L.}~\bibnamefont {Wang}},\
  }\bibfield  {title} {\bibinfo {title} {Differentiable learning of quantum
  circuit born machines},\ }\href {https://doi.org/10.1103/PhysRevA.98.062324}
  {\bibfield  {journal} {\bibinfo  {journal} {Phys. Rev. A}\ }\textbf {\bibinfo
  {volume} {98}},\ \bibinfo {pages} {062324} (\bibinfo {year}
  {2018})}\BibitemShut {NoStop}%
\bibitem [{\citenamefont {Lemke}\ \emph {et~al.}(2015)\citenamefont {Lemke},
  \citenamefont {Budka},\ and\ \citenamefont {Gabrys}}]{Lemke2015}%
  \BibitemOpen
  \bibfield  {author} {\bibinfo {author} {\bibfnamefont {C.}~\bibnamefont
  {Lemke}}, \bibinfo {author} {\bibfnamefont {M.}~\bibnamefont {Budka}},\ and\
  \bibinfo {author} {\bibfnamefont {B.}~\bibnamefont {Gabrys}},\ }\bibfield
  {title} {\bibinfo {title} {Metalearning: a survey of trends and
  technologies},\ }\href {https://doi.org/10.1007/s10462-013-9406-y} {\bibfield
   {journal} {\bibinfo  {journal} {Artificial Intelligence Review}\ }\textbf
  {\bibinfo {volume} {44}},\ \bibinfo {pages} {117} (\bibinfo {year}
  {2015})}\BibitemShut {NoStop}%
\bibitem [{\citenamefont {{Hospedales}}\ \emph {et~al.}(2020)\citenamefont
  {{Hospedales}}, \citenamefont {{Antoniou}}, \citenamefont {{Micaelli}},\ and\
  \citenamefont {{Storkey}}}]{2020arXiv200405439H}%
  \BibitemOpen
  \bibfield  {author} {\bibinfo {author} {\bibfnamefont {T.}~\bibnamefont
  {{Hospedales}}}, \bibinfo {author} {\bibfnamefont {A.}~\bibnamefont
  {{Antoniou}}}, \bibinfo {author} {\bibfnamefont {P.}~\bibnamefont
  {{Micaelli}}},\ and\ \bibinfo {author} {\bibfnamefont {A.}~\bibnamefont
  {{Storkey}}},\ }\bibfield  {title} {\bibinfo {title} {{Meta-Learning in
  Neural Networks: A Survey}},\ }\href {http://arxiv.org/abs/2004.05439}
  {\bibfield  {journal} {\bibinfo  {journal} {arXiv:2004.05439}\ } (\bibinfo
  {year} {2020})}\BibitemShut {NoStop}%
\bibitem [{\citenamefont {Mari}\ \emph {et~al.}(2021)\citenamefont {Mari},
  \citenamefont {Bromley},\ and\ \citenamefont
  {Killoran}}]{PhysRevA.103.012405}%
  \BibitemOpen
  \bibfield  {author} {\bibinfo {author} {\bibfnamefont {A.}~\bibnamefont
  {Mari}}, \bibinfo {author} {\bibfnamefont {T.~R.}\ \bibnamefont {Bromley}},\
  and\ \bibinfo {author} {\bibfnamefont {N.}~\bibnamefont {Killoran}},\
  }\bibfield  {title} {\bibinfo {title} {Estimating the gradient and
  higher-order derivatives on quantum hardware},\ }\href
  {https://doi.org/10.1103/PhysRevA.103.012405} {\bibfield  {journal} {\bibinfo
   {journal} {Phys. Rev. A}\ }\textbf {\bibinfo {volume} {103}},\ \bibinfo
  {pages} {012405} (\bibinfo {year} {2021})}\BibitemShut {NoStop}%
\bibitem [{\citenamefont {Pagano}\ \emph {et~al.}(2020)\citenamefont {Pagano},
  \citenamefont {Bapat}, \citenamefont {Becker}, \citenamefont {Collins},
  \citenamefont {De}, \citenamefont {Hess}, \citenamefont {Kaplan},
  \citenamefont {Kyprianidis}, \citenamefont {Tan}, \citenamefont {Baldwin},
  \citenamefont {Brady}, \citenamefont {Deshpande}, \citenamefont {Liu},
  \citenamefont {Jordan}, \citenamefont {Gorshkov},\ and\ \citenamefont
  {Monroe}}]{Pagano25396}%
  \BibitemOpen
  \bibfield  {author} {\bibinfo {author} {\bibfnamefont {G.}~\bibnamefont
  {Pagano}}, \bibinfo {author} {\bibfnamefont {A.}~\bibnamefont {Bapat}},
  \bibinfo {author} {\bibfnamefont {P.}~\bibnamefont {Becker}}, \bibinfo
  {author} {\bibfnamefont {K.~S.}\ \bibnamefont {Collins}}, \bibinfo {author}
  {\bibfnamefont {A.}~\bibnamefont {De}}, \bibinfo {author} {\bibfnamefont
  {P.~W.}\ \bibnamefont {Hess}}, \bibinfo {author} {\bibfnamefont {H.~B.}\
  \bibnamefont {Kaplan}}, \bibinfo {author} {\bibfnamefont {A.}~\bibnamefont
  {Kyprianidis}}, \bibinfo {author} {\bibfnamefont {W.~L.}\ \bibnamefont
  {Tan}}, \bibinfo {author} {\bibfnamefont {C.}~\bibnamefont {Baldwin}},
  \bibinfo {author} {\bibfnamefont {L.~T.}\ \bibnamefont {Brady}}, \bibinfo
  {author} {\bibfnamefont {A.}~\bibnamefont {Deshpande}}, \bibinfo {author}
  {\bibfnamefont {F.}~\bibnamefont {Liu}}, \bibinfo {author} {\bibfnamefont
  {S.}~\bibnamefont {Jordan}}, \bibinfo {author} {\bibfnamefont {A.~V.}\
  \bibnamefont {Gorshkov}},\ and\ \bibinfo {author} {\bibfnamefont
  {C.}~\bibnamefont {Monroe}},\ }\bibfield  {title} {\bibinfo {title} {Quantum
  approximate optimization of the long-range ising model with a trapped-ion
  quantum simulator},\ }\href {https://doi.org/10.1073/pnas.2006373117}
  {\bibfield  {journal} {\bibinfo  {journal} {Proc. Natl. Acad. Sci.}\ }\textbf
  {\bibinfo {volume} {117}},\ \bibinfo {pages} {25396} (\bibinfo {year}
  {2020})}\BibitemShut {NoStop}%
\bibitem [{\citenamefont {Bergholm}\ \emph {et~al.}(2018)\citenamefont
  {Bergholm}, \citenamefont {Izaac}, \citenamefont {Schuld}, \citenamefont
  {Gogolin}, \citenamefont {Alam}, \citenamefont {Ahmed}, \citenamefont
  {Arrazola}, \citenamefont {Blank}, \citenamefont {Delgado}, \citenamefont
  {Jahangiri} \emph {et~al.}}]{bergholm2018pennylane}%
  \BibitemOpen
  \bibfield  {author} {\bibinfo {author} {\bibfnamefont {V.}~\bibnamefont
  {Bergholm}}, \bibinfo {author} {\bibfnamefont {J.}~\bibnamefont {Izaac}},
  \bibinfo {author} {\bibfnamefont {M.}~\bibnamefont {Schuld}}, \bibinfo
  {author} {\bibfnamefont {C.}~\bibnamefont {Gogolin}}, \bibinfo {author}
  {\bibfnamefont {M.~S.}\ \bibnamefont {Alam}}, \bibinfo {author}
  {\bibfnamefont {S.}~\bibnamefont {Ahmed}}, \bibinfo {author} {\bibfnamefont
  {J.~M.}\ \bibnamefont {Arrazola}}, \bibinfo {author} {\bibfnamefont
  {C.}~\bibnamefont {Blank}}, \bibinfo {author} {\bibfnamefont
  {A.}~\bibnamefont {Delgado}}, \bibinfo {author} {\bibfnamefont
  {S.}~\bibnamefont {Jahangiri}}, \emph {et~al.},\ }\bibfield  {title}
  {\bibinfo {title} {Pennylane: Automatic differentiation of hybrid
  quantum-classical computations},\ }\href {http://arxiv.org/abs/1811.04968}
  {\bibfield  {journal} {\bibinfo  {journal} {arXiv:1811.04968}\ } (\bibinfo
  {year} {2018})}\BibitemShut {NoStop}%
\bibitem [{\citenamefont {Broughton}\ \emph {et~al.}(2020)\citenamefont
  {Broughton}, \citenamefont {Verdon}, \citenamefont {McCourt}, \citenamefont
  {Martinez}, \citenamefont {Yoo}, \citenamefont {Isakov}, \citenamefont
  {Massey}, \citenamefont {Niu}, \citenamefont {Halavati}, \citenamefont
  {Peters} \emph {et~al.}}]{broughton2020tensorflow}%
  \BibitemOpen
  \bibfield  {author} {\bibinfo {author} {\bibfnamefont {M.}~\bibnamefont
  {Broughton}}, \bibinfo {author} {\bibfnamefont {G.}~\bibnamefont {Verdon}},
  \bibinfo {author} {\bibfnamefont {T.}~\bibnamefont {McCourt}}, \bibinfo
  {author} {\bibfnamefont {A.~J.}\ \bibnamefont {Martinez}}, \bibinfo {author}
  {\bibfnamefont {J.~H.}\ \bibnamefont {Yoo}}, \bibinfo {author} {\bibfnamefont
  {S.~V.}\ \bibnamefont {Isakov}}, \bibinfo {author} {\bibfnamefont
  {P.}~\bibnamefont {Massey}}, \bibinfo {author} {\bibfnamefont {M.~Y.}\
  \bibnamefont {Niu}}, \bibinfo {author} {\bibfnamefont {R.}~\bibnamefont
  {Halavati}}, \bibinfo {author} {\bibfnamefont {E.}~\bibnamefont {Peters}},
  \emph {et~al.},\ }\bibfield  {title} {\bibinfo {title} {Tensorflow quantum: A
  software framework for quantum machine learning},\ }\href
  {http://arxiv.org/abs/2003.02989} {\bibfield  {journal} {\bibinfo  {journal}
  {arXiv:2003.02989}\ } (\bibinfo {year} {2020})}\BibitemShut {NoStop}%
\bibitem [{\citenamefont {Edward~Farhi}(2014)}]{Farhi2014}%
  \BibitemOpen
  \bibfield  {author} {\bibinfo {author} {\bibfnamefont {S.~G.}\ \bibnamefont
  {Edward~Farhi}, \bibfnamefont {Jeffrey~Goldstone}},\ }\bibfield  {title}
  {\bibinfo {title} {A quantum approximate optimization algorithm},\ }\href
  {http://arxiv.org/abs/1411.4028} {\bibfield  {journal} {\bibinfo  {journal}
  {arXiv:1411.4028}\ } (\bibinfo {year} {2014})}\BibitemShut {NoStop}%
\bibitem [{\citenamefont {Brandao}\ \emph {et~al.}(2018)\citenamefont
  {Brandao}, \citenamefont {Broughton}, \citenamefont {Farhi}, \citenamefont
  {Gutmann},\ and\ \citenamefont {Neven}}]{brandao2018fixed}%
  \BibitemOpen
  \bibfield  {author} {\bibinfo {author} {\bibfnamefont {F.~G.}\ \bibnamefont
  {Brandao}}, \bibinfo {author} {\bibfnamefont {M.}~\bibnamefont {Broughton}},
  \bibinfo {author} {\bibfnamefont {E.}~\bibnamefont {Farhi}}, \bibinfo
  {author} {\bibfnamefont {S.}~\bibnamefont {Gutmann}},\ and\ \bibinfo {author}
  {\bibfnamefont {H.}~\bibnamefont {Neven}},\ }\bibfield  {title} {\bibinfo
  {title} {For fixed control parameters the quantum approximate optimization
  algorithm's objective function value concentrates for typical instances},\
  }\href {http://arxiv.org/abs/1812.04170} {\bibfield  {journal} {\bibinfo
  {journal} {arXiv:1812.04170}\ } (\bibinfo {year} {2018})}\BibitemShut
  {NoStop}%
\bibitem [{\citenamefont {Crooks}(2018)}]{crooks2018performance}%
  \BibitemOpen
  \bibfield  {author} {\bibinfo {author} {\bibfnamefont {G.~E.}\ \bibnamefont
  {Crooks}},\ }\bibfield  {title} {\bibinfo {title} {Performance of the quantum
  approximate optimization algorithm on the maximum cut problem},\ }\href
  {http://arxiv.org/abs/1811.08419} {\bibfield  {journal} {\bibinfo  {journal}
  {arXiv:1811.08419}\ } (\bibinfo {year} {2018})}\BibitemShut {NoStop}%
\bibitem [{\citenamefont {Willsch}\ \emph {et~al.}(2020)\citenamefont
  {Willsch}, \citenamefont {Willsch}, \citenamefont {Jin}, \citenamefont
  {De~Raedt},\ and\ \citenamefont {Michielsen}}]{Willsch2020}%
  \BibitemOpen
  \bibfield  {author} {\bibinfo {author} {\bibfnamefont {M.}~\bibnamefont
  {Willsch}}, \bibinfo {author} {\bibfnamefont {D.}~\bibnamefont {Willsch}},
  \bibinfo {author} {\bibfnamefont {F.}~\bibnamefont {Jin}}, \bibinfo {author}
  {\bibfnamefont {H.}~\bibnamefont {De~Raedt}},\ and\ \bibinfo {author}
  {\bibfnamefont {K.}~\bibnamefont {Michielsen}},\ }\bibfield  {title}
  {\bibinfo {title} {Benchmarking the quantum approximate optimization
  algorithm},\ }\href {https://doi.org/10.1007/s11128-020-02692-8} {\bibfield
  {journal} {\bibinfo  {journal} {Quantum Inf. Proc.}\ }\textbf {\bibinfo
  {volume} {19}},\ \bibinfo {pages} {197} (\bibinfo {year} {2020})}\BibitemShut
  {NoStop}%
\bibitem [{\citenamefont {Kandala}\ \emph {et~al.}(2017)\citenamefont
  {Kandala}, \citenamefont {Mezzacapo}, \citenamefont {Temme}, \citenamefont
  {Takita}, \citenamefont {Brink}, \citenamefont {Chow},\ and\ \citenamefont
  {Gambetta}}]{Kandala2017}%
  \BibitemOpen
  \bibfield  {author} {\bibinfo {author} {\bibfnamefont {A.}~\bibnamefont
  {Kandala}}, \bibinfo {author} {\bibfnamefont {A.}~\bibnamefont {Mezzacapo}},
  \bibinfo {author} {\bibfnamefont {K.}~\bibnamefont {Temme}}, \bibinfo
  {author} {\bibfnamefont {M.}~\bibnamefont {Takita}}, \bibinfo {author}
  {\bibfnamefont {M.}~\bibnamefont {Brink}}, \bibinfo {author} {\bibfnamefont
  {J.~M.}\ \bibnamefont {Chow}},\ and\ \bibinfo {author} {\bibfnamefont
  {J.~M.}\ \bibnamefont {Gambetta}},\ }\bibfield  {title} {\bibinfo {title}
  {Hardware-efficient variational quantum eigensolver for small molecules and
  quantum magnets},\ }\href {https://doi.org/10.1038/nature23879} {\bibfield
  {journal} {\bibinfo  {journal} {Nature}\ }\textbf {\bibinfo {volume} {549}},\
  \bibinfo {pages} {242} (\bibinfo {year} {2017})}\BibitemShut {NoStop}%
\bibitem [{\citenamefont {Dallaire-Demers}\ \emph {et~al.}(2019)\citenamefont
  {Dallaire-Demers}, \citenamefont {Romero}, \citenamefont {Veis},
  \citenamefont {Sim},\ and\ \citenamefont
  {Aspuru-Guzik}}]{Dallaire_Demers_2019}%
  \BibitemOpen
  \bibfield  {author} {\bibinfo {author} {\bibfnamefont {P.-L.}\ \bibnamefont
  {Dallaire-Demers}}, \bibinfo {author} {\bibfnamefont {J.}~\bibnamefont
  {Romero}}, \bibinfo {author} {\bibfnamefont {L.}~\bibnamefont {Veis}},
  \bibinfo {author} {\bibfnamefont {S.}~\bibnamefont {Sim}},\ and\ \bibinfo
  {author} {\bibfnamefont {A.}~\bibnamefont {Aspuru-Guzik}},\ }\bibfield
  {title} {\bibinfo {title} {Low-depth circuit ansatz for preparing correlated
  fermionic states on a quantum computer},\ }\href
  {https://doi.org/10.1088/2058-9565/ab3951} {\bibfield  {journal} {\bibinfo
  {journal} {Quantum Sci. Technol.}\ }\textbf {\bibinfo {volume} {4}},\
  \bibinfo {pages} {045005} (\bibinfo {year} {2019})}\BibitemShut {NoStop}%
\bibitem [{\citenamefont {Peruzzo}\ \emph {et~al.}(2014)\citenamefont
  {Peruzzo}, \citenamefont {McClean}, \citenamefont {Shadbolt}, \citenamefont
  {Yung}, \citenamefont {Zhou}, \citenamefont {Love}, \citenamefont
  {Aspuru-Guzik},\ and\ \citenamefont {O'Brien}}]{Peruzzo2014}%
  \BibitemOpen
  \bibfield  {author} {\bibinfo {author} {\bibfnamefont {A.}~\bibnamefont
  {Peruzzo}}, \bibinfo {author} {\bibfnamefont {J.}~\bibnamefont {McClean}},
  \bibinfo {author} {\bibfnamefont {P.}~\bibnamefont {Shadbolt}}, \bibinfo
  {author} {\bibfnamefont {M.-H.}\ \bibnamefont {Yung}}, \bibinfo {author}
  {\bibfnamefont {X.-Q.}\ \bibnamefont {Zhou}}, \bibinfo {author}
  {\bibfnamefont {P.~J.}\ \bibnamefont {Love}}, \bibinfo {author}
  {\bibfnamefont {A.}~\bibnamefont {Aspuru-Guzik}},\ and\ \bibinfo {author}
  {\bibfnamefont {J.~L.}\ \bibnamefont {O'Brien}},\ }\bibfield  {title}
  {\bibinfo {title} {A variational eigenvalue solver on a photonic quantum
  processor},\ }\href {https://doi.org/10.1038/ncomms5213} {\bibfield
  {journal} {\bibinfo  {journal} {Nat. Commun.}\ }\textbf {\bibinfo {volume}
  {5}},\ \bibinfo {pages} {4213 EP } (\bibinfo {year} {2014})}\BibitemShut
  {NoStop}%
\bibitem [{Note1()}]{Note1}%
  \BibitemOpen
  \bibinfo {note} {Https://www.orquestra.io/}\BibitemShut {NoStop}%
\bibitem [{\citenamefont {Lacroix}\ \emph {et~al.}(2020)\citenamefont
  {Lacroix}, \citenamefont {Hellings}, \citenamefont {Andersen}, \citenamefont
  {Di~Paolo}, \citenamefont {Remm}, \citenamefont {Lazar}, \citenamefont
  {Krinner}, \citenamefont {Norris}, \citenamefont {Gabureac}, \citenamefont
  {Heinsoo}, \citenamefont {Blais}, \citenamefont {Eichler},\ and\
  \citenamefont {Wallraff}}]{PRXQuantum.1.020304}%
  \BibitemOpen
  \bibfield  {author} {\bibinfo {author} {\bibfnamefont {N.}~\bibnamefont
  {Lacroix}}, \bibinfo {author} {\bibfnamefont {C.}~\bibnamefont {Hellings}},
  \bibinfo {author} {\bibfnamefont {C.~K.}\ \bibnamefont {Andersen}}, \bibinfo
  {author} {\bibfnamefont {A.}~\bibnamefont {Di~Paolo}}, \bibinfo {author}
  {\bibfnamefont {A.}~\bibnamefont {Remm}}, \bibinfo {author} {\bibfnamefont
  {S.}~\bibnamefont {Lazar}}, \bibinfo {author} {\bibfnamefont
  {S.}~\bibnamefont {Krinner}}, \bibinfo {author} {\bibfnamefont {G.~J.}\
  \bibnamefont {Norris}}, \bibinfo {author} {\bibfnamefont {M.}~\bibnamefont
  {Gabureac}}, \bibinfo {author} {\bibfnamefont {J.}~\bibnamefont {Heinsoo}},
  \bibinfo {author} {\bibfnamefont {A.}~\bibnamefont {Blais}}, \bibinfo
  {author} {\bibfnamefont {C.}~\bibnamefont {Eichler}},\ and\ \bibinfo {author}
  {\bibfnamefont {A.}~\bibnamefont {Wallraff}},\ }\bibfield  {title} {\bibinfo
  {title} {Improving the performance of deep quantum optimization algorithms
  with continuous gate sets},\ }\href
  {https://doi.org/10.1103/PRXQuantum.1.020304} {\bibfield  {journal} {\bibinfo
   {journal} {PRX Quantum}\ }\textbf {\bibinfo {volume} {1}},\ \bibinfo {pages}
  {110304} (\bibinfo {year} {2020})}\BibitemShut {NoStop}%
\bibitem [{\citenamefont {Harrigan}\ \emph {et~al.}(2021)\citenamefont
  {Harrigan}, \citenamefont {Sung}, \citenamefont {Neeley}, \citenamefont
  {Satzinger}, \citenamefont {Arute}, \citenamefont {Arya}, \citenamefont
  {Atalaya}, \citenamefont {Bardin}, \citenamefont {Barends}, \citenamefont
  {Boixo}, \citenamefont {Broughton}, \citenamefont {Buckley}, \citenamefont
  {Buell}, \citenamefont {Burkett}, \citenamefont {Bushnell}, \citenamefont
  {Chen}, \citenamefont {Chen}, \citenamefont {Chiaro}, \citenamefont
  {Collins}, \citenamefont {Courtney}, \citenamefont {Demura}, \citenamefont
  {Dunsworth}, \citenamefont {Eppens}, \citenamefont {Fowler}, \citenamefont
  {Foxen}, \citenamefont {Gidney}, \citenamefont {Giustina}, \citenamefont
  {Graff}, \citenamefont {Habegger}, \citenamefont {Ho}, \citenamefont {Hong},
  \citenamefont {Huang}, \citenamefont {Ioffe}, \citenamefont {Isakov},
  \citenamefont {Jeffrey}, \citenamefont {Jiang}, \citenamefont {Jones},
  \citenamefont {Kafri}, \citenamefont {Kechedzhi}, \citenamefont {Kelly},
  \citenamefont {Kim}, \citenamefont {Klimov}, \citenamefont {Korotkov},
  \citenamefont {Kostritsa}, \citenamefont {Landhuis}, \citenamefont {Laptev},
  \citenamefont {Lindmark}, \citenamefont {Leib}, \citenamefont {Martin},
  \citenamefont {Martinis}, \citenamefont {McClean}, \citenamefont {McEwen},
  \citenamefont {Megrant}, \citenamefont {Mi}, \citenamefont {Mohseni},
  \citenamefont {Mruczkiewicz}, \citenamefont {Mutus}, \citenamefont {Naaman},
  \citenamefont {Neill}, \citenamefont {Neukart}, \citenamefont {Niu},
  \citenamefont {O'Brien}, \citenamefont {O'Gorman}, \citenamefont {Ostby},
  \citenamefont {Petukhov}, \citenamefont {Putterman}, \citenamefont
  {Quintana}, \citenamefont {Roushan}, \citenamefont {Rubin}, \citenamefont
  {Sank}, \citenamefont {Skolik}, \citenamefont {Smelyanskiy}, \citenamefont
  {Strain}, \citenamefont {Streif}, \citenamefont {Szalay}, \citenamefont
  {Vainsencher}, \citenamefont {White}, \citenamefont {Yao}, \citenamefont
  {Yeh}, \citenamefont {Zalcman}, \citenamefont {Zhou}, \citenamefont {Neven},
  \citenamefont {Bacon}, \citenamefont {Lucero}, \citenamefont {Farhi},\ and\
  \citenamefont {Babbush}}]{Harrigan2021}%
  \BibitemOpen
  \bibfield  {author} {\bibinfo {author} {\bibfnamefont {M.~P.}\ \bibnamefont
  {Harrigan}}, \bibinfo {author} {\bibfnamefont {K.~J.}\ \bibnamefont {Sung}},
  \bibinfo {author} {\bibfnamefont {M.}~\bibnamefont {Neeley}}, \bibinfo
  {author} {\bibfnamefont {K.~J.}\ \bibnamefont {Satzinger}}, \bibinfo {author}
  {\bibfnamefont {F.}~\bibnamefont {Arute}}, \bibinfo {author} {\bibfnamefont
  {K.}~\bibnamefont {Arya}}, \bibinfo {author} {\bibfnamefont {J.}~\bibnamefont
  {Atalaya}}, \bibinfo {author} {\bibfnamefont {J.~C.}\ \bibnamefont {Bardin}},
  \bibinfo {author} {\bibfnamefont {R.}~\bibnamefont {Barends}}, \bibinfo
  {author} {\bibfnamefont {S.}~\bibnamefont {Boixo}}, \bibinfo {author}
  {\bibfnamefont {M.}~\bibnamefont {Broughton}}, \bibinfo {author}
  {\bibfnamefont {B.~B.}\ \bibnamefont {Buckley}}, \bibinfo {author}
  {\bibfnamefont {D.~A.}\ \bibnamefont {Buell}}, \bibinfo {author}
  {\bibfnamefont {B.}~\bibnamefont {Burkett}}, \bibinfo {author} {\bibfnamefont
  {N.}~\bibnamefont {Bushnell}}, \bibinfo {author} {\bibfnamefont
  {Y.}~\bibnamefont {Chen}}, \bibinfo {author} {\bibfnamefont {Z.}~\bibnamefont
  {Chen}}, \bibinfo {author} {\bibfnamefont {B.}~\bibnamefont {Chiaro}},
  \bibinfo {author} {\bibfnamefont {R.}~\bibnamefont {Collins}}, \bibinfo
  {author} {\bibfnamefont {W.}~\bibnamefont {Courtney}}, \bibinfo {author}
  {\bibfnamefont {S.}~\bibnamefont {Demura}}, \bibinfo {author} {\bibfnamefont
  {A.}~\bibnamefont {Dunsworth}}, \bibinfo {author} {\bibfnamefont
  {D.}~\bibnamefont {Eppens}}, \bibinfo {author} {\bibfnamefont
  {A.}~\bibnamefont {Fowler}}, \bibinfo {author} {\bibfnamefont
  {B.}~\bibnamefont {Foxen}}, \bibinfo {author} {\bibfnamefont
  {C.}~\bibnamefont {Gidney}}, \bibinfo {author} {\bibfnamefont
  {M.}~\bibnamefont {Giustina}}, \bibinfo {author} {\bibfnamefont
  {R.}~\bibnamefont {Graff}}, \bibinfo {author} {\bibfnamefont
  {S.}~\bibnamefont {Habegger}}, \bibinfo {author} {\bibfnamefont
  {A.}~\bibnamefont {Ho}}, \bibinfo {author} {\bibfnamefont {S.}~\bibnamefont
  {Hong}}, \bibinfo {author} {\bibfnamefont {T.}~\bibnamefont {Huang}},
  \bibinfo {author} {\bibfnamefont {L.~B.}\ \bibnamefont {Ioffe}}, \bibinfo
  {author} {\bibfnamefont {S.~V.}\ \bibnamefont {Isakov}}, \bibinfo {author}
  {\bibfnamefont {E.}~\bibnamefont {Jeffrey}}, \bibinfo {author} {\bibfnamefont
  {Z.}~\bibnamefont {Jiang}}, \bibinfo {author} {\bibfnamefont
  {C.}~\bibnamefont {Jones}}, \bibinfo {author} {\bibfnamefont
  {D.}~\bibnamefont {Kafri}}, \bibinfo {author} {\bibfnamefont
  {K.}~\bibnamefont {Kechedzhi}}, \bibinfo {author} {\bibfnamefont
  {J.}~\bibnamefont {Kelly}}, \bibinfo {author} {\bibfnamefont
  {S.}~\bibnamefont {Kim}}, \bibinfo {author} {\bibfnamefont {P.~V.}\
  \bibnamefont {Klimov}}, \bibinfo {author} {\bibfnamefont {A.~N.}\
  \bibnamefont {Korotkov}}, \bibinfo {author} {\bibfnamefont {F.}~\bibnamefont
  {Kostritsa}}, \bibinfo {author} {\bibfnamefont {D.}~\bibnamefont {Landhuis}},
  \bibinfo {author} {\bibfnamefont {P.}~\bibnamefont {Laptev}}, \bibinfo
  {author} {\bibfnamefont {M.}~\bibnamefont {Lindmark}}, \bibinfo {author}
  {\bibfnamefont {M.}~\bibnamefont {Leib}}, \bibinfo {author} {\bibfnamefont
  {O.}~\bibnamefont {Martin}}, \bibinfo {author} {\bibfnamefont {J.~M.}\
  \bibnamefont {Martinis}}, \bibinfo {author} {\bibfnamefont {J.~R.}\
  \bibnamefont {McClean}}, \bibinfo {author} {\bibfnamefont {M.}~\bibnamefont
  {McEwen}}, \bibinfo {author} {\bibfnamefont {A.}~\bibnamefont {Megrant}},
  \bibinfo {author} {\bibfnamefont {X.}~\bibnamefont {Mi}}, \bibinfo {author}
  {\bibfnamefont {M.}~\bibnamefont {Mohseni}}, \bibinfo {author} {\bibfnamefont
  {W.}~\bibnamefont {Mruczkiewicz}}, \bibinfo {author} {\bibfnamefont
  {J.}~\bibnamefont {Mutus}}, \bibinfo {author} {\bibfnamefont
  {O.}~\bibnamefont {Naaman}}, \bibinfo {author} {\bibfnamefont
  {C.}~\bibnamefont {Neill}}, \bibinfo {author} {\bibfnamefont
  {F.}~\bibnamefont {Neukart}}, \bibinfo {author} {\bibfnamefont {M.~Y.}\
  \bibnamefont {Niu}}, \bibinfo {author} {\bibfnamefont {T.~E.}\ \bibnamefont
  {O'Brien}}, \bibinfo {author} {\bibfnamefont {B.}~\bibnamefont {O'Gorman}},
  \bibinfo {author} {\bibfnamefont {E.}~\bibnamefont {Ostby}}, \bibinfo
  {author} {\bibfnamefont {A.}~\bibnamefont {Petukhov}}, \bibinfo {author}
  {\bibfnamefont {H.}~\bibnamefont {Putterman}}, \bibinfo {author}
  {\bibfnamefont {C.}~\bibnamefont {Quintana}}, \bibinfo {author}
  {\bibfnamefont {P.}~\bibnamefont {Roushan}}, \bibinfo {author} {\bibfnamefont
  {N.~C.}\ \bibnamefont {Rubin}}, \bibinfo {author} {\bibfnamefont
  {D.}~\bibnamefont {Sank}}, \bibinfo {author} {\bibfnamefont {A.}~\bibnamefont
  {Skolik}}, \bibinfo {author} {\bibfnamefont {V.}~\bibnamefont {Smelyanskiy}},
  \bibinfo {author} {\bibfnamefont {D.}~\bibnamefont {Strain}}, \bibinfo
  {author} {\bibfnamefont {M.}~\bibnamefont {Streif}}, \bibinfo {author}
  {\bibfnamefont {M.}~\bibnamefont {Szalay}}, \bibinfo {author} {\bibfnamefont
  {A.}~\bibnamefont {Vainsencher}}, \bibinfo {author} {\bibfnamefont
  {T.}~\bibnamefont {White}}, \bibinfo {author} {\bibfnamefont {Z.~J.}\
  \bibnamefont {Yao}}, \bibinfo {author} {\bibfnamefont {P.}~\bibnamefont
  {Yeh}}, \bibinfo {author} {\bibfnamefont {A.}~\bibnamefont {Zalcman}},
  \bibinfo {author} {\bibfnamefont {L.}~\bibnamefont {Zhou}}, \bibinfo {author}
  {\bibfnamefont {H.}~\bibnamefont {Neven}}, \bibinfo {author} {\bibfnamefont
  {D.}~\bibnamefont {Bacon}}, \bibinfo {author} {\bibfnamefont
  {E.}~\bibnamefont {Lucero}}, \bibinfo {author} {\bibfnamefont
  {E.}~\bibnamefont {Farhi}},\ and\ \bibinfo {author} {\bibfnamefont
  {R.}~\bibnamefont {Babbush}},\ }\bibfield  {title} {\bibinfo {title} {Quantum
  approximate optimization of non-planar graph problems on a planar
  superconducting processor},\ }\href
  {https://doi.org/10.1038/s41567-020-01105-y} {\bibfield  {journal} {\bibinfo
  {journal} {Nat. Phys.}\ }\textbf {\bibinfo {volume} {17}},\ \bibinfo {pages}
  {332} (\bibinfo {year} {2021})}\BibitemShut {NoStop}%
\bibitem [{\citenamefont {Wiersema}\ \emph {et~al.}(2020)\citenamefont
  {Wiersema}, \citenamefont {Zhou}, \citenamefont {de~Sereville}, \citenamefont
  {Carrasquilla}, \citenamefont {Kim},\ and\ \citenamefont
  {Yuen}}]{wiersema2020exploring}%
  \BibitemOpen
  \bibfield  {author} {\bibinfo {author} {\bibfnamefont {R.}~\bibnamefont
  {Wiersema}}, \bibinfo {author} {\bibfnamefont {C.}~\bibnamefont {Zhou}},
  \bibinfo {author} {\bibfnamefont {Y.}~\bibnamefont {de~Sereville}}, \bibinfo
  {author} {\bibfnamefont {J.~F.}\ \bibnamefont {Carrasquilla}}, \bibinfo
  {author} {\bibfnamefont {Y.~B.}\ \bibnamefont {Kim}},\ and\ \bibinfo {author}
  {\bibfnamefont {H.}~\bibnamefont {Yuen}},\ }\bibfield  {title} {\bibinfo
  {title} {Exploring entanglement and optimization within the hamiltonian
  variational ansatz},\ }\href {https://doi.org/10.1103/PRXQuantum.1.020319}
  {\bibfield  {journal} {\bibinfo  {journal} {PRX Quantum}\ }\textbf {\bibinfo
  {volume} {1}},\ \bibinfo {pages} {020319} (\bibinfo {year}
  {2020})}\BibitemShut {NoStop}%
\bibitem [{\citenamefont {Li}\ \emph {et~al.}(2020)\citenamefont {Li},
  \citenamefont {Fan}, \citenamefont {Coram}, \citenamefont {Riley},\ and\
  \citenamefont {Leichenauer}}]{PhysRevResearch.2.023074}%
  \BibitemOpen
  \bibfield  {author} {\bibinfo {author} {\bibfnamefont {L.}~\bibnamefont
  {Li}}, \bibinfo {author} {\bibfnamefont {M.}~\bibnamefont {Fan}}, \bibinfo
  {author} {\bibfnamefont {M.}~\bibnamefont {Coram}}, \bibinfo {author}
  {\bibfnamefont {P.}~\bibnamefont {Riley}},\ and\ \bibinfo {author}
  {\bibfnamefont {S.}~\bibnamefont {Leichenauer}},\ }\bibfield  {title}
  {\bibinfo {title} {Quantum optimization with a novel gibbs objective function
  and ansatz architecture search},\ }\href
  {https://doi.org/10.1103/PhysRevResearch.2.023074} {\bibfield  {journal}
  {\bibinfo  {journal} {Phys. Rev. Research}\ }\textbf {\bibinfo {volume}
  {2}},\ \bibinfo {pages} {023074} (\bibinfo {year} {2020})}\BibitemShut
  {NoStop}%
\bibitem [{\citenamefont {Dagotto}(1994)}]{Dagotto1994}%
  \BibitemOpen
  \bibfield  {author} {\bibinfo {author} {\bibfnamefont {E.}~\bibnamefont
  {Dagotto}},\ }\bibfield  {title} {\bibinfo {title} {{Correlated electrons in
  high-temperature superconductors}},\ }\href
  {https://doi.org/10.1103/RevModPhys.66.763} {\bibfield  {journal} {\bibinfo
  {journal} {Rev. Mod. Phys.}\ }\textbf {\bibinfo {volume} {66}},\ \bibinfo
  {pages} {763} (\bibinfo {year} {1994})}\BibitemShut {NoStop}%
\bibitem [{\citenamefont {LeBlanc}\ \emph {et~al.}(2015)\citenamefont
  {LeBlanc}, \citenamefont {Antipov}, \citenamefont {Becca}, \citenamefont
  {Bulik}, \citenamefont {Chan}, \citenamefont {Chung}, \citenamefont {Deng},
  \citenamefont {Ferrero}, \citenamefont {Henderson}, \citenamefont
  {Jim\'enez-Hoyos}, \citenamefont {Kozik}, \citenamefont {Liu}, \citenamefont
  {Millis}, \citenamefont {Prokof'ev}, \citenamefont {Qin}, \citenamefont
  {Scuseria}, \citenamefont {Shi}, \citenamefont {Svistunov}, \citenamefont
  {Tocchio}, \citenamefont {Tupitsyn}, \citenamefont {White}, \citenamefont
  {Zhang}, \citenamefont {Zheng}, \citenamefont {Zhu},\ and\ \citenamefont
  {Gull}}]{LeBlanc2015}%
  \BibitemOpen
  \bibfield  {author} {\bibinfo {author} {\bibfnamefont {J.~P.~F.}\
  \bibnamefont {LeBlanc}}, \bibinfo {author} {\bibfnamefont {A.~E.}\
  \bibnamefont {Antipov}}, \bibinfo {author} {\bibfnamefont {F.}~\bibnamefont
  {Becca}}, \bibinfo {author} {\bibfnamefont {I.~W.}\ \bibnamefont {Bulik}},
  \bibinfo {author} {\bibfnamefont {G.~K.-L.}\ \bibnamefont {Chan}}, \bibinfo
  {author} {\bibfnamefont {C.-M.}\ \bibnamefont {Chung}}, \bibinfo {author}
  {\bibfnamefont {Y.}~\bibnamefont {Deng}}, \bibinfo {author} {\bibfnamefont
  {M.}~\bibnamefont {Ferrero}}, \bibinfo {author} {\bibfnamefont {T.~M.}\
  \bibnamefont {Henderson}}, \bibinfo {author} {\bibfnamefont {C.~A.}\
  \bibnamefont {Jim\'enez-Hoyos}}, \bibinfo {author} {\bibfnamefont
  {E.}~\bibnamefont {Kozik}}, \bibinfo {author} {\bibfnamefont {X.-W.}\
  \bibnamefont {Liu}}, \bibinfo {author} {\bibfnamefont {A.~J.}\ \bibnamefont
  {Millis}}, \bibinfo {author} {\bibfnamefont {N.~V.}\ \bibnamefont
  {Prokof'ev}}, \bibinfo {author} {\bibfnamefont {M.}~\bibnamefont {Qin}},
  \bibinfo {author} {\bibfnamefont {G.~E.}\ \bibnamefont {Scuseria}}, \bibinfo
  {author} {\bibfnamefont {H.}~\bibnamefont {Shi}}, \bibinfo {author}
  {\bibfnamefont {B.~V.}\ \bibnamefont {Svistunov}}, \bibinfo {author}
  {\bibfnamefont {L.~F.}\ \bibnamefont {Tocchio}}, \bibinfo {author}
  {\bibfnamefont {I.~S.}\ \bibnamefont {Tupitsyn}}, \bibinfo {author}
  {\bibfnamefont {S.~R.}\ \bibnamefont {White}}, \bibinfo {author}
  {\bibfnamefont {S.}~\bibnamefont {Zhang}}, \bibinfo {author} {\bibfnamefont
  {B.-X.}\ \bibnamefont {Zheng}}, \bibinfo {author} {\bibfnamefont
  {Z.}~\bibnamefont {Zhu}},\ and\ \bibinfo {author} {\bibfnamefont
  {E.}~\bibnamefont {Gull}},\ }\bibfield  {title} {\bibinfo {title} {Solutions
  of the two-dimensional hubbard model: Benchmarks and results from a wide
  range of numerical algorithms},\ }\href
  {https://doi.org/10.1103/PhysRevX.5.041041} {\bibfield  {journal} {\bibinfo
  {journal} {Phys. Rev. X}\ }\textbf {\bibinfo {volume} {5}},\ \bibinfo {pages}
  {041041} (\bibinfo {year} {2015})}\BibitemShut {NoStop}%
\bibitem [{\citenamefont {Cade}\ \emph {et~al.}(2020)\citenamefont {Cade},
  \citenamefont {Mineh}, \citenamefont {Montanaro},\ and\ \citenamefont
  {Stanisic}}]{Cade2020}%
  \BibitemOpen
  \bibfield  {author} {\bibinfo {author} {\bibfnamefont {C.}~\bibnamefont
  {Cade}}, \bibinfo {author} {\bibfnamefont {L.}~\bibnamefont {Mineh}},
  \bibinfo {author} {\bibfnamefont {A.}~\bibnamefont {Montanaro}},\ and\
  \bibinfo {author} {\bibfnamefont {S.}~\bibnamefont {Stanisic}},\ }\bibfield
  {title} {\bibinfo {title} {{Strategies for solving the Fermi-Hubbard model on
  near-term quantum computers}},\ }\href
  {https://doi.org/10.1103/PhysRevB.102.235122} {\bibfield  {journal} {\bibinfo
   {journal} {Phys. Rev. B}\ }\textbf {\bibinfo {volume} {102}},\ \bibinfo
  {pages} {235122} (\bibinfo {year} {2020})},\ \Eprint
  {https://arxiv.org/abs/1912.06007} {1912.06007} \BibitemShut {NoStop}%
\bibitem [{\citenamefont {Lieb}\ and\ \citenamefont {Wu}(2003)}]{lieb2003one}%
  \BibitemOpen
  \bibfield  {author} {\bibinfo {author} {\bibfnamefont {E.~H.}\ \bibnamefont
  {Lieb}}\ and\ \bibinfo {author} {\bibfnamefont {F.}~\bibnamefont {Wu}},\
  }\bibfield  {title} {\bibinfo {title} {The one-dimensional hubbard model: a
  reminiscence},\ }\href
  {https://doi.org/https://doi.org/10.1016/S0378-4371(02)01785-5} {\bibfield
  {journal} {\bibinfo  {journal} {Physica A: Statistical Mechanics and its
  Applications}\ }\textbf {\bibinfo {volume} {321}},\ \bibinfo {pages} {1}
  (\bibinfo {year} {2003})}\BibitemShut {NoStop}%
\bibitem [{\citenamefont {Dallaire-Demers}\ \emph {et~al.}(2020)\citenamefont
  {Dallaire-Demers}, \citenamefont {Stech{\l}y}, \citenamefont {Gonthier},
  \citenamefont {Bashige}, \citenamefont {Romero},\ and\ \citenamefont
  {Cao}}]{dallairedemers2020application}%
  \BibitemOpen
  \bibfield  {author} {\bibinfo {author} {\bibfnamefont {P.-L.}\ \bibnamefont
  {Dallaire-Demers}}, \bibinfo {author} {\bibfnamefont {M.}~\bibnamefont
  {Stech{\l}y}}, \bibinfo {author} {\bibfnamefont {J.~F.}\ \bibnamefont
  {Gonthier}}, \bibinfo {author} {\bibfnamefont {N.~T.}\ \bibnamefont
  {Bashige}}, \bibinfo {author} {\bibfnamefont {J.}~\bibnamefont {Romero}},\
  and\ \bibinfo {author} {\bibfnamefont {Y.}~\bibnamefont {Cao}},\ }\bibfield
  {title} {\bibinfo {title} {{An application benchmark for fermionic quantum
  simulations}},\ }\href {http://arxiv.org/abs/2003.01862} {\bibfield
  {journal} {\bibinfo  {journal} {arXiv:2003.01862}\ } (\bibinfo {year}
  {2020})}\BibitemShut {NoStop}%
\bibitem [{\citenamefont {Sim}\ \emph {et~al.}(2021)\citenamefont {Sim},
  \citenamefont {Romero}, \citenamefont {Gonthier},\ and\ \citenamefont
  {Kunitsa}}]{sim2020adaptive}%
  \BibitemOpen
  \bibfield  {author} {\bibinfo {author} {\bibfnamefont {S.}~\bibnamefont
  {Sim}}, \bibinfo {author} {\bibfnamefont {J.}~\bibnamefont {Romero}},
  \bibinfo {author} {\bibfnamefont {J.~F.}\ \bibnamefont {Gonthier}},\ and\
  \bibinfo {author} {\bibfnamefont {A.~A.}\ \bibnamefont {Kunitsa}},\
  }\bibfield  {title} {\bibinfo {title} {Adaptive pruning-based optimization of
  parameterized quantum circuits},\ }\href
  {https://doi.org/10.1088/2058-9565/abe107} {\bibfield  {journal} {\bibinfo
  {journal} {Quantum Sci. Technol.}\ }\textbf {\bibinfo {volume} {6}},\
  \bibinfo {pages} {025019} (\bibinfo {year} {2021})}\BibitemShut {NoStop}%
\bibitem [{\citenamefont {Campos}\ \emph {et~al.}(2021)\citenamefont {Campos},
  \citenamefont {Nasrallah},\ and\ \citenamefont
  {Biamonte}}]{campos2020abrupt}%
  \BibitemOpen
  \bibfield  {author} {\bibinfo {author} {\bibfnamefont {E.}~\bibnamefont
  {Campos}}, \bibinfo {author} {\bibfnamefont {A.}~\bibnamefont {Nasrallah}},\
  and\ \bibinfo {author} {\bibfnamefont {J.}~\bibnamefont {Biamonte}},\
  }\bibfield  {title} {\bibinfo {title} {Abrupt transitions in variational
  quantum circuit training},\ }\href
  {https://doi.org/10.1103/PhysRevA.103.032607} {\bibfield  {journal} {\bibinfo
   {journal} {Phys. Rev. A}\ }\textbf {\bibinfo {volume} {103}},\ \bibinfo
  {pages} {032607} (\bibinfo {year} {2021})}\BibitemShut {NoStop}%
\bibitem [{\citenamefont {Flennerhag}\ \emph {et~al.}(2019)\citenamefont
  {Flennerhag}, \citenamefont {Rusu}, \citenamefont {Pascanu}, \citenamefont
  {Visin}, \citenamefont {Yin},\ and\ \citenamefont
  {Hadsell}}]{flennerhag2019meta}%
  \BibitemOpen
  \bibfield  {author} {\bibinfo {author} {\bibfnamefont {S.}~\bibnamefont
  {Flennerhag}}, \bibinfo {author} {\bibfnamefont {A.~A.}\ \bibnamefont
  {Rusu}}, \bibinfo {author} {\bibfnamefont {R.}~\bibnamefont {Pascanu}},
  \bibinfo {author} {\bibfnamefont {F.}~\bibnamefont {Visin}}, \bibinfo
  {author} {\bibfnamefont {H.}~\bibnamefont {Yin}},\ and\ \bibinfo {author}
  {\bibfnamefont {R.}~\bibnamefont {Hadsell}},\ }\bibfield  {title} {\bibinfo
  {title} {Meta-learning with warped gradient descent},\ }\href
  {http://arxiv.org/abs/1909.00025} {\bibfield  {journal} {\bibinfo  {journal}
  {arXiv:1909.00025}\ } (\bibinfo {year} {2019})}\BibitemShut {NoStop}%
\bibitem [{Note2()}]{Note2}%
  \BibitemOpen
  \bibinfo {note} {We refer to a layer of nearest-neighboring two-qubit gates
  that start with acting on the first and second qubits as an even layer, and
  one that start with acting on the second and third qubits as an odd
  layer.}\BibitemShut {Stop}%
\bibitem [{\citenamefont {Kingma}\ and\ \citenamefont
  {Ba}(2014)}]{kingma2014adam}%
  \BibitemOpen
  \bibfield  {author} {\bibinfo {author} {\bibfnamefont {D.~P.}\ \bibnamefont
  {Kingma}}\ and\ \bibinfo {author} {\bibfnamefont {J.}~\bibnamefont {Ba}},\
  }\bibfield  {title} {\bibinfo {title} {Adam: A method for stochastic
  optimization},\ }\href {http://arxiv.org/abs/1412.6980} {\bibfield  {journal}
  {\bibinfo  {journal} {arXiv:1412.6980}\ } (\bibinfo {year}
  {2014})}\BibitemShut {NoStop}%
\bibitem [{\citenamefont {Liu}\ \emph {et~al.}(2020)\citenamefont {Liu},
  \citenamefont {Jiang}, \citenamefont {He}, \citenamefont {Chen},
  \citenamefont {Liu}, \citenamefont {Gao},\ and\ \citenamefont
  {Han}}]{DBLP:conf/iclr/LiuJHCLG020}%
  \BibitemOpen
  \bibfield  {author} {\bibinfo {author} {\bibfnamefont {L.}~\bibnamefont
  {Liu}}, \bibinfo {author} {\bibfnamefont {H.}~\bibnamefont {Jiang}}, \bibinfo
  {author} {\bibfnamefont {P.}~\bibnamefont {He}}, \bibinfo {author}
  {\bibfnamefont {W.}~\bibnamefont {Chen}}, \bibinfo {author} {\bibfnamefont
  {X.}~\bibnamefont {Liu}}, \bibinfo {author} {\bibfnamefont {J.}~\bibnamefont
  {Gao}},\ and\ \bibinfo {author} {\bibfnamefont {J.}~\bibnamefont {Han}},\
  }\bibfield  {title} {\bibinfo {title} {On the variance of the adaptive
  learning rate and beyond},\ }in\ \href
  {https://openreview.net/forum?id=rkgz2aEKDr} {\emph {\bibinfo {booktitle}
  {8th International Conference on Learning Representations, {ICLR} 2020, Addis
  Ababa, Ethiopia}}}\ (\bibinfo {year} {2020})\BibitemShut {NoStop}%
\bibitem [{\citenamefont {Albash}\ and\ \citenamefont
  {Lidar}(2018)}]{RevModPhys.90.015002}%
  \BibitemOpen
  \bibfield  {author} {\bibinfo {author} {\bibfnamefont {T.}~\bibnamefont
  {Albash}}\ and\ \bibinfo {author} {\bibfnamefont {D.~A.}\ \bibnamefont
  {Lidar}},\ }\bibfield  {title} {\bibinfo {title} {Adiabatic quantum
  computation},\ }\href {https://doi.org/10.1103/RevModPhys.90.015002}
  {\bibfield  {journal} {\bibinfo  {journal} {Rev. Mod. Phys.}\ }\textbf
  {\bibinfo {volume} {90}},\ \bibinfo {pages} {015002} (\bibinfo {year}
  {2018})}\BibitemShut {NoStop}%
\end{thebibliography}
\end{document}